\documentclass[]{emulateapj}

\usepackage{natbib}
\usepackage{amsmath}
\usepackage{epsfig}
\usepackage{graphicx}
\usepackage{amssymb}
\usepackage{color}
\usepackage{epstopdf}
\usepackage{rotating}
\newcommand{\be}{\begin{equation}}
\newcommand{\ee}{\end{equation}}
\newcommand{\ba}{\begin{eqnarray}}
\newcommand{\ea}{\end{eqnarray}}

\newcommand{\lya}{\mbox{Ly-$\alpha\ $}}
\newcommand{\lyn}{\mbox{Ly-$n\ $}}
\newcommand{\jcap}{\ref@jnl{J. Cosmology Astropart. Phys.}}

\begin{document}

\title{Intensity Mapping of Lyman-alpha Emission During the Epoch of Reionization}

\author{Marta B. Silva$^{1,2}$, Mario G. Santos$^1$, Yan Gong$^2$, Asantha Cooray$^2$ and James Bock$^{3,4}$}

\affil{$^1$CENTRA, Instituto Superior T\'ecnico, Technical University of Lisbon, Lisboa 1049-001, Portugal}
\affil{$^2$Department of Physics \& Astronomy, University of California, Irvine, CA 92697}
\affil{$^3$Department of Physics, Mathematics and Astronomy, California Institute of Technology, Pasadena, CA 91125, USA}
\affil{$^4$Jet Propulsion Laboratory (JPL), National Aeronautics and Space Administration (NASA), Pasadena, CA 91109, USA}

\begin{abstract}
We calculate the absolute intensity and anisotropies of the Lyman-$\alpha$ radiation field present during the epoch of reionization. We consider emission from both galaxies and the intergalactic medium (IGM) and take into account the main contributions to the production of Lyman-$\alpha$ photons:  recombinations, collisions, continuum emission from galaxies and scattering of Lyman-n photons in the IGM. We find that the emission from individual galaxies dominates over the IGM with a total Lyman-$\alpha$ intensity (times frequency) of about $(1.43-3.57)\times10^{-8} {\rm erg\ s^{-1}\ cm^{-2}\ sr^{-1}}$ at a redshift of 7. This intensity level is low so it is unlikely that the Lyman-$\alpha$ background during reionization can be established by an experiment aiming at an absolute background light measurement. Instead we consider Lyman-$\alpha$ intensity mapping with the aim of measuring the anisotropy power spectrum which has rms fluctuations at the level of $1 \times 10^{-16} {\rm [erg\ s^{-1}\ cm^{-2}\ sr^{-1}]
^2}$ at a few Mpc scales. These anisotropies could be measured with a spectrometer at near-IR wavelengths from 0.9 to 1.4 $\mu$m with fields in the order of 0.5 to 1 sq. degrees. We recommend that existing ground-based programs using narrow band filters also pursue intensity fluctuations to study statistics on the spatial distribution of faint Lyman-$\alpha$ emitters.  We also discuss the cross-correlation signal with 21 cm experiments that probe HI in the IGM during reionization. A dedicated sub-orbital or space-based Lyman-$\alpha$ intensity mapping experiment could provide a viable complimentary approach to probe reionization, when compared to 21 cm experiments, and is likely within experimental reach.
\end{abstract}


\maketitle

\section{Introduction}

The epoch of reionization (EoR) is a crucial stage in the history of galaxy formation, signaling the birth of the first luminous objects, during which the universe went from completely neutral to almost completely ionized \citep{2001PhR...349..125B}. This phase has been largely unexplored so far, although current observations suggest it was reasonably extended \citep{2011ApJS..192...18K,2006AJ....132..117F} and a wide variety of observational avenues are being explored to probe it. In particular the 21-cm line of neutral hydrogen is now understood to be a promising tool to study reionization
and to understand the formation and evolution of galaxies during that epoch (see e.g. \citealt{2006PhR...433..181F}). It is also now becoming clear that we need complimentary data in order 
to obtain  extra insight into the sources of reionization. Such complimentary data could also aid in the interpretation of the HI signal by allowing ways to pursue cross-correlations and
providing ways to reduce systematics and foregrounds encountered in 21-cm observations.

Recently, intensity mapping of other atomic and molecular lines at high redshifts, in particular CO and CII \citep{2012ApJ...745...49G, 2011ApJ...728L..46G, 2011ApJ...741...70L, 2010JCAP...11..016V}, 
has been proposed as a probe of reionization. In this work we study the viability of also using intensity mapping of the Lyman-$\alpha$ (Ly-$\alpha$) line as an additional probe.
For this study we include several \lya emission mechanisms involving both individual sources of emission such as galaxies
and the emission and scattering associated with the intergalactic medium (IGM).

We consider both the integrated intensity and anisotropies of the \lya line and suggest the latter as a new probe of reionization. In particular we suggest that it will be possible to measure the amplitude of the 
\lya intensity fluctuations with a narrow-band spectrometer either from the ground with a suppression of atmospheric lines or from the orbital/sub-orbital platform.

The \lya line, corresponding to transitions between the second and first energy level of the hydrogen atom, has a rest wavelength of approximately $\lambda_{\rm Ly\alpha}=1216\ {\rm \mathring{A}}$. The signal present during reionization is observable in near-IR wavelengths today. Existing imaging observations made with narrow-band filters on 10m class telescopes focus on individual galaxy detections
and are limited to a handful of narrow atmospheric windows at near-IR wavelengths. Given the strength of the line, it has now been seen in galaxies at
$z\approx 6.98$ \citep{2006Natur.443..186I}, $z\approx 8.2$ \citep{2009Natur.461.1258S} and $z\approx 8.6$ \citep{2010Natur.467..940L}, reaching well into the epoch of reionization.

Deep narrow-band surveys of high redshift \lya emitters have led to detections of a sufficient number of galaxies at redshifts $5.7$, $6.6$, $7.0$ and $7.7$ to allow constraints on the bright-end of the \lya luminosity function (LF) and its redshift evolution (e.g. \citealt{2008ApJS..176..301O,2010ApJ...722..803O,2005PASJ...57..165T,2006Natur.443..186I,2011arXiv1112.3997S}). 
Observations of the \lya LF indicate a decrease in the \lya intensity from redshift $5.7$ to $7.0$. This would require a strong evolution of the \lya emitters population, which is not predicted by most recent galaxy evolution models \citep{2010ApJ...722..803O,2011arXiv1112.3997S}, or could be explained as the result of an increase in the fraction of IGM neutral hydrogen that would absorb or scatter \lya photons from the observed galaxies \citep{2000ApJ...537L...5H,2008ApJ...677...12O}. 

The scattering of \lya photons by neutral hydrogen in the ISM (interstellar medium) and the IGM is expected to disperse the photons in both frequency and direction \citep{2004MNRAS.349.1137S}. 
Such scattering could considerably decrease the \lya intensity per frequency bin from an individual galaxy, making the detection of most of the 
high redshift galaxies impossible with current instruments. Exact calculations related to scattering are a difficult problem to solve analytically and in simulations the scattering problem requires ray tracing of photons through the 
neutral medium in a simulation box \citep{2010ApJ...716..574Z}. While scattering makes individual galaxies dimmer,  intensity mapping of the \lya line at high redshifts can be an improvement over the
usual experiments that make detections of \lya emission from point sources and are only sensible to the strongest \lya emitters. These are likely to be some of the brightest star-forming galaxies, however, any dust that is present in such galaxies, especially during the late stages of reionization, is likely to suppress the \lya line. An experiment targeting the integrated emission will be able to measure all the sources of \lya photons in a large region and will be sensitive to the extended, low surface brightness \lya emission that is now known to
generally form around star-forming regions (e.g., \citealt{2011ApJ...736..160S,2011ApJ...739...62Z}). The anisotropy power spectrum of \lya intensity then would be a probe of the \lya halos around star-forming galaxies
present during reionization. The cross-correlation with the 21-cm data could provide a direct test on the presence of neutral hydrogen in the extended \lya halo.

The paper is organized as follows: in the next section we estimate the contribution to the \lya emission from galaxies. In section \ref{sec:IGM} we analyze the contributions to the \lya emission from the IGM. In section \ref{sec:sim} we calculate the intensity of the \lya signal as well as its power spectrum using a modified version of the code SimFast21 \citep{2010MNRAS.406.2421S,2011A&A...527A..93S}. In section \ref{sec:cross_correlation} we discuss the correlation of \lya intensity maps with the 21 cm signal and finally in section 6 we comment on the experimental feasibility of measuring the \lya intensity power spectrum.

\section{Lyman-$\alpha$ emission from Galaxies}
\label{sec:GAL}

The observed \lya flux is mainly the result of line emission from hydrogen recombinations and collisional excitations in the interstellar clouds or in the IGM powered respectively by UV emission or UV and X-ray emission from galaxies. High energy photons emitted by stars ionize hydrogen that then recombines to emit a rich spectrum of lines including a \lya photon \citep{1996ApJ...468..462G,2006ApJ...646..703F}. Moreover, the electron ejected during this ionization heats the ISM or the IGM, increasing the probability of \lya photon emission caused by collisional excitation \citep{1996ApJ...468..462G,2008ApJ...672...48C}. 
There is also a small contribution to the lyman alpha flux originated in the 
continuum emission from stars between the \lya line and the Lyman-limit \citep{2007ApJ...670..912C,2005ApJ...626....1B}  plus \lya
from continuum free-free or free-bound emission as well as 2-photon emission during recombinations. This continuum will also make contributions to a given observation from lower redshifts besides the "\lya" redshift \citep{2012ApJ...756...92C} which will confuse the \lya signal. However, due to the smoothness of that continuum across frequency, we expect it should be possible to remove this contribution, for instance, by fitting a smooth polynomial in frequency for each pixel.

Another source of \lya emission in the universe is cooling of gas that has suffered in-fall into a dark matter halo potential well. Several studies show that much of this cooling is made in the form of \lya emission \citep{2000ApJ...537L...5H,2001ApJ...562..605F,2006ApJ...649...14D,2006ApJ...649...37D,2010MNRAS.402.1449D,2011MNRAS.413L..33L}. Cold gas is used by galaxies as fuel to form stars so there is a relation between the star formation rate (SFR) of a galaxy and the \lya flux emitted as gas cools in that galaxy. 

Since emission of \lya radiation is closely connected with the star formation, the contribution from the several mechanisms by which \lya radiation is emitted in galaxies and in the IGM can be related to the SFR of individual galaxies or galaxy samples. In order to calculate the emission of \lya radiation from the IGM during the EoR we also need to know the ionized fraction of hydrogen as well as the temperature of the gas in the IGM. Unfortunately both these quantities are poorly constrained at $z\ge 6$ \citep{2011ApJS..192...16L,2010ApJ...723..869O,2011arXiv1111.6386Z}. Since hydrogen ionization should be a consequence of stellar ionization/X-ray emission, we can in principle estimate it by following the SFR history and making sure that the resulting evolution of hydrogen ionized fraction is consistent with current constraints on the CMB optical depth.

In order to obtain the SFR of galaxies at the high redshifts during the epoch of reionization we make use of parametrizations that reproduce a correct reionization history. Our parametrizations are non linear in a similar way to the relations found in the \citet{2011MNRAS.413..101G} and in the \citet{2007MNRAS.375....2D} galaxies catalogs derived respectively from the high resolution Millennium II \citep{2009MNRAS.398.1150B} and Millennium I \citep{2005Natur.435..629S} simulations. Such relations, when available from observations, make an improvement on the models instead of relying purely on 
theoretical calculations and semi-numerical simulations to predict all of the observations \citep{2007ApJ...669..663M,2010MNRAS.406.2421S}. 

There are additional sources of radiation contributing to the \lya emission, such as a strong non-local sources of ionizing photons as expected from quasars, which would emit a large amount of energy in X-ray photons that would be able to ionize several neutral atoms giving origin to a locally strong \lya emission from recombinations. However, since the number of quasars is very small compared to the number of normal galaxies at the redshifts we are considering, we will neglect their contribution in the following calculations. 

We encourage future works on \lya intensity to see if the shape of the power spectrum and other statistics can be used to choose between reionization histories that involve both galaxies and quasars.

In the following sub-sections we discuss in more detail the four processes for \lya emission from galaxies: recombinations, excitations/relaxations, gas cooling, and photon emission from continuum processes.

\subsection{Lyman-$\alpha$ emission from hydrogen recombinations}
\label{GAL:rec}

Assuming ionizing equilibrium, the number of recombinations in galaxies are expected to match the number of ionizing photons that are absorbed in the galaxy and does not escape into the IGM.
Depending on the temperature and density of the gas, a fraction of the radiation due to these recombinations is emitted in the \lya line.

In the interstellar gas, most of the neutral hydrogen is in dense clouds with column densities greater than $3\times10^{18}$ cm$^{-2}$. These clouds are optically thick to \lya radiation 
and Lyman photons are scattered in the galaxy several times before escaping into the IGM. Such multiple scatterings increase the probability of absorption. 
Assuming that these clouds are spherical and that the gas temperature is of the order of $10^4$ K, \citet{1996ApJ...468..462G} used atomic physics to study the probability 
of the \lya emission per hydrogen recombination. They estimated that a fraction $f_{rec}\approx 66\%$ of the hydrogen recombinations would result in the emission of a \lya photon and that 
most of the other recombinations would result in two-photon emission. These fractions should change with the temperature and the shape of the cloud, but such variations are expected to be small. 
Other calculations yield fractions between $62$\% and $68$\% according to the conditions in the cloud. In this paper we have chosen to use a value of $f_{rec}= 66\%$ since 
the overall uncertainty on this number is lower than the uncertainty on the number of hydrogen recombinations. 

The absorption of \lya photons by dust is difficult to estimate and changes from galaxy to galaxy, \citet{1996ApJ...468..462G} estimated that for a cloud with a column density $N\sim10^{19}$ cm$^{-2}$, the dust in the galaxy absorbs a fraction $f_{\rm dust}\approx 4$\% of the emitted \lya photons before they reach the galaxy virial radius however recent observations of high redshift galaxies indicate a much higher $f_{\rm dust}$. In this study we will use a redshift parameterization for the fraction of \lya photons that are not absorbed by dust $f_{\rm Ly\alpha}=1-f_{\rm dust}$ that is double the value predicted by the study made by \citet{2011ApJ...730....8H}:
\be
f_{\rm Ly\alpha}(z)=C_{\rm dust}\times10^{-3}(1+z)^{\xi},
\label{eq:f_Lyalpha}
\ee
where $C_{\rm dust}=3.34$ and $\xi=2.57$. The \citet{2011ApJ...730....8H} parameterization was made so that $f_{\rm Ly\alpha}$ gives the difference between observed \lya luminosities and \lya luminosities scaled from star formation rates assuming that the \lya alpha photons emitted in galaxies are only originated in recombinations. The high redshift observations used to estimate $f_{\rm Ly\alpha}$ are only of massive stars while the bulk of \lya emission is originated in the low mass stars that cannot be
detected by current surveys. According to several studies \citep{2011MNRAS.415.3666F}, $f_{\rm Ly\alpha}$ decreases with halo mass, so it is possible that it is being underestimated in \citet{2011ApJ...730....8H} which is why we decided to use a higher $f_{\rm Ly\alpha}$.
Our results can however be easily scaled to other $f_{\rm dust}$ evolutions. 

The number of \lya photons emitted in a galaxy per second, $\dot{N}_{\rm Ly\alpha}$, that reach its virial radius is therefore given by
\be
\dot{N}_{\rm Ly\alpha} = A_{He} f_{\rm rec}\times f_{\rm Ly\alpha}\times (1-f_{\rm esc})\times \dot{N}_{\rm ion},
\ee
where $A_{\rm He}=\frac{4-4Y_p}{4-3Y_p}$ accounts for the fraction of photons that go into the ionization of helium ($Y_p$ is the mass fraction of helium), $\dot{N}_{\rm ion}$ is the rate of ionizing photons emitted by the stars in the galaxy and $f_{\rm esc}$ is the fraction of ionizing photons that escape the galaxy into the IGM.

The ionizing photon escape fraction depends on conditions inside each galaxy and is difficult to estimate, especially at high redshifts. The precise determination of its value is one of the 
major goals of future observations of high redshift galaxies at $z\gtrsim 7$.
This parameter can be measured from deep imaging observations or can be estimated from the equivalent widths of the hydrogen and helium balmer lines.
The ionizing photon escape fraction dependence with the galaxy mass and the star formation rate, as a function of redshift,
 has been estimated using simulations that make several assumptions about the intensity of this radiation and its absorption in the interstellar medium. However, for the halo virial mass range, $10^8 {\rm M_{\odot}}$ to $10^{13} {\rm M_{\odot}}$, and during the broad redshift range related to the epoch of reionization,
there are no simulations that cover the full parameter space. Moreover the limited simulations that exist do not always agree with each other \citep{2008ApJ...672..765G,2009ApJ...693..984W,2003MNRAS.342.1215F,2007ApJ...668...62S,2012ApJ...746..125H}. \cite{2010ApJ...710.1239R} computed the escape fraction of UV radiation for the redshift interval $z=4$ to $z=10$ and for halos of masses from $10^{7.8}$ to $10^{11.5}$ M$_{\sun}$ 
using a high-resolution set of galaxies. Their simulations cover most of the parameter space needed for reionization related calculations and their
 escape fraction parameterization is compatible with most of the current observational results. Thus, we use it for our calculations here.

According to \cite{2010ApJ...710.1239R} simulations, the escape fraction of ionizing radiation can be parameterized as:
\be
f_{\rm esc}(M,z)=\exp\left[-\alpha(z) M^{\beta (z)}\right],
\ee
where $M$ is the halo mass, $\alpha$ and $\beta$ are functions of redshift (Table~\ref{tab:fesc}).
\begin{table}
\centering                         
\begin{tabular}{l  c c c}        
\hline\hline                 
z & $\alpha$ & $\beta$ & $f_{\rm esc}(M=10^{\rm 10}$ $M_{\odot})$\\    
\hline                        
   10.4 & $2.78\times10^{\rm -2}$ & $0.105$ & $0.732$ \\
   8.2 & $1.30\times10^{\rm -2}$ & $0.179$  & $0.449$\\
   6.7 & $5.18\times10^{\rm -3}$ & $0.244$ & $0.240$\\
   5.7 & $3.42\times10^{\rm -3} $ & $0.262$ & $0.240$\\
\hline                                  
\end{tabular}
\caption{Fits to the escape fraction of UV radiation from galaxies as a function of redshift (based on Razoumov \& Sommer-Larsen 2010).}
\label{tab:fesc}     
\end{table}
The number of ionizing photons emitted by the stars in a galaxy depends on its star formation rate, metallicity and the stellar initial mass function (IMF).
Making reasonable assumptions for these quantities we will now estimate $\dot{N}_{\rm ion}$. Since this UV emission is dominated by massive, short lived stars, 
we can assume that the intensity of ionizing photons emitted by a galaxy is proportional to its star formation rate.
In terms of the star formation rate in one galaxy,
\be 
\dot{N}_{\rm ion}=Q_{\rm ion}\times {\rm SFR},
\label{eq:Nion}
\ee 
where $Q_{\rm ion}$ is the average number of ionizing photons emitted per solar mass of star formation. This can be calculated through:
\be
Q_{\rm ion}=\frac{\int^{\rm M_{max}} _{\rm M_{min}} \Psi(M) Q_{\rm \star}(M) t_{\rm \star}(M) dM}{\int^{\rm M_{max}} _{\rm M_{min}} \Psi(M) M dM},
\label{eq:Qion}
\ee
where  $\Psi(M)=KM^{-\alpha}$ is the stellar IMF, $K$ is a constant normalization factor and $\alpha$ is the slope of the IMF. In our calculation we used a Salpeter IMF, with $\alpha=2.35$. $t_{\rm \star}(M)$ is the star lifetime and $Q_{\rm \star}(M)$ its number of ionizing photons emitted per unit time.
The values of $Q_{\rm \star}$ and $t_{\rm \star}$ were calculated with the ionizing fluxes obtained by \citet{2002A&A...382...28S} using realistic models of stellar populations and non-LTE atmospheric models, appropriated for POP II stars with a $Z_{\rm \star}=0.02Z_{\rm \odot}$ metallicity.

Assuming that ionizing photons are only emitted by massive OB stars sets a low mass effective limit for the mass of stars contributing to the UV radiation field of a galaxy. 
This limit is a necessary condition for the star to be able to produce a significant number of ionizing photons. For the stellar population used for this work
we take $M_{\rm min}\approx 7$ M$_{\odot}$ \citep{2002A&A...382...28S,2011arXiv1108.3334S}. The integration upper limit is taken to be $M_{\rm max}=150$ M$_{\odot}$. In this paper we calculated Q$_{ion}$ using the parameterization values published in \citet{2002A&A...382...28S}. The number of ionizing photons per 
second emitted by a star as a function of its mass is given by:
\ba
\log_{10}[Q_{\star}/{\rm s}^{-1}]=&27.80&+30.68x-14.80x^{\rm 2}\\ \nonumber     
&+&2.5x^3\ {\rm for}\ 7\, {\rm M_{\rm \odot}}<M_{\rm \star}<150\, {\rm M_{\rm \odot}}
\label{eq:photons_second}
\ea
where $x=\log_{\rm 10}(M_{\rm \star}/M_{\rm \odot})$ and the star's lifetime in years is given by:
\be
\log_{\rm 10}[t_\star/{\rm yr}]=9.59-2.79x+0.63x^{\rm 2}.
\label{eq:star_life_year}
\ee
The use of these parameters results in $Q_{\rm {ion}}\approx 5.38\times10^{\rm 60}$ M$_{\odot}^{\rm -1}$. In \citet{2011arXiv1108.3334S} it has been suggested the use of a different model for stellar atmosphere and evolution (R. S. Sutherland $\&$ J. M. Shull, unpublished) which yields $Q_{\rm ion}\approx 3.97\times10^{\rm 60}$ M$_{\rm \odot}^{\rm -1}$. This may imply that the stellar emissivity we calculated is an overestimation and that consequently our \lya flux powered by stellar emission may be overestimated by about 35\%. This is comparable to other large uncertainties, such as the ones in the parameters $f_{\rm esc}$ and $f_{\rm dust}$. 
The \lya luminosity is calculated assuming that the \lya photons are emitted at the \lya rest frequency, $\nu_{\rm 0}=2.47\times10^{\rm 15}$ Hz with an energy of $E_{\rm Ly\alpha}=1.637\times10^{\rm -11}$ erg. 
To proceed, we will assume that the SFR for a given galaxy is only a function of redshift and the mass of the dark halo associated with that galaxy. The \lya luminosity due to recombinations in the interstellar 
medium, $L^{\rm GAL}_{\rm rec}$, can then be parameterized as a function of halo mass and redshift as
\ba
&&L^{\rm GAL}_{\rm rec}(M,z)=E_{\rm Ly\alpha}\dot{N}_{\rm Ly\alpha}\\ \nonumber
&\approx&  1.55 \times 10^{\rm 42} \left[1-f_{\rm esc}(M,z)\right]f_{\rm Ly\alpha}(z)\frac{{\rm SFR}(M,z)}{{\rm M}_\odot\; {\rm yr}^{-1}}{\rm erg}\ s^{-1}.
\ea

\subsection{Lyman-$\alpha$ emission from excitations/relaxations}

The kinetic energy of the electron ejected during the hydrogen ionization heats the gas and assuming thermal equilibrium this heat is emitted as radiation.
Using atomic physics, \citet{1996ApJ...468..462G} estimated that for a cloud with an hydrogen column density of $\approx 10^{\rm 19}$ cm$^{-2}$, the energy emitted in the form of \lya photons is about $60$\% for ionizing photons with energy $E_{\nu_{\rm lim}}<E_{\nu}<4 E_{\nu_{\rm lim}}$ and $\approx 50$\% for photons with energy $E_{\nu}>4 E_{\nu_{\rm lim}}$, where $E_{\nu_{\rm lim}}=13.6$ eV is the Rydberg energy. The remaining of the energy is emitted in other lines.

Using the spectral energy distribution (SED) of galaxies with a metallicity $Z=0.02 Z_{\odot}$ from the code of \cite{2005MNRAS.362..799M} we estimated that the average ionizing photon energy is $E_{\nu}=21.4$ eV and that more than 99\% of the photons have an energy lower than $4E_{\nu_{\rm lim}}$.  
According to the \citet{1996ApJ...468..462G} calculation, the fraction of energy of the UV photon that is emitted as Lyman alpha radiation due to the collisional excitations/relaxations is given by:
\be
E_{\rm exc}/E_{\nu} \sim 0.08+0.1\left(1- \frac{2\nu_{\rm lim}}{\nu} \right) \sim 0.1
\ee 
For a cloud with the properties considered here this yields an energy in \lya per ionizing photon of $E_{\rm exc}\approx 2.14$ eV or $3.43\times10^{-12}$ erg. This results in an average of $0.16$ \lya photons per ionizing photon.

Finally, the \lya luminosity due to excitations in the ISM, $L^{\rm GAL}_{\rm exc}$, is then:
\ba
L^{\rm  GAL}_{\rm exc}(M,z)&=&[1-f_{\rm esc}(M,z)]f_{\rm Ly\alpha}(z)A_{\rm He}\\ \nonumber
&\times& \dot{N}_{\rm ion}E_{\rm exc}\\ \nonumber
\approx 4.03\times&10^{41}&\left[1-f_{\rm esc}(M,z)\right]f_{\rm Ly\alpha}(z)\frac{{\rm SFR}(M,z)}{{\rm M}_\odot \, {\rm yr}^{-1}}{\rm erg}\ s^{-1},
\ea
where again it is assumed to be a function of the star formation rate.

\subsection{Lyman-$\alpha$ emission from gas cooling}

During the formation of galaxies, gas from the IGM falls into potential wells composed mainly by dark matter which collapsed under its own gravity. 
The increase in the gas density leads to a high rate of atomic collisions that heats the gas to a high temperature. According to the study of \citet{2001ApJ...562..605F} most of the gas 
in potential wells that collapses under its own gravity never reaches its virial temperature and so a large fraction of the potential energy is released by line emission induced by collisions and excitations from gas with temperatures $T_K<2 \times 10^4$ K. At this temperature approximately $50\%$ of the energy is emitted in \lya alone.

From \citet{2001ApJ...562..605F} we can relate the luminosity at the \lya frequency due to the cooling in galaxies to their baryonic cold mass, $M^{\rm bar}_{\rm cool}$, using:
\be
\log_{10}\left( L^{\rm GAL}_{\rm cool}\right)=1.52\log_{10}(M^{\rm bar}_{\rm cool})+26.32,
\label{eq:ecoll}
\ee
where both the luminosity and the mass are in solar units.
To relate this baryonic cold mass to a quantity we can use in our models, we used the relation between cold baryonic mass and the halo mass from the galaxies in the \cite{2011MNRAS.413..101G} catalog. From the equation above, we can then obtain an expression for the luminosity, which can be fitted by:
\ba
L^{\rm GAL}_{\rm cool}(M) &\approx& 1.69 \times 10^{35} f_{\rm Ly\alpha}(z)\left(1+\frac{M}{10^{\rm 8}}\right)\\ \nonumber
&\times& \left(1+\frac{M}{2\times10^{\rm 10}}\right)^{2.1}\left(1+\frac{M}{3\times10^{\rm 11}}\right)^{-3} {\rm erg}\ s^{-1},
\label{eq:ecoll2}
\ea
with $M$ in units of $M_{\rm \odot}$.
The relation between the cold gas mass and the mass of the halo
shows very little evolution with redshift during reionization.  Thus we expect the
relation in equation~\ref{eq:ecoll2} to only depend on redshift due to the redshift evolution of $f_{\rm Ly\alpha}$.

\subsection{ Contributions from continuum emission}

Continuum emission can also contribute to the \lya observations. These include stellar emission, free-free emission, free-bound emission and two photon emission. Photons emitted with frequencies close to the \lyn lines should scatter within the ISM and eventually get re-emitted out of the galaxy as \lya photons. Otherwise they will escape the ISM before redshifting into one of the \lyn lines and being reabsorbed by a hydrogen atom.

The fraction of photons that scatter in the galaxy can be estimated from the intrinsic width of the Lyman alpha line which has $\approx 4 {\rm \mathring{A}}$ \citep{2012arXiv1206.4028J}.
We calculated the stellar contribution assuming an emission spectrum for stars with a metallicity of $Z_{\star}=0.02Z_{\odot}$ estimated with the code from \cite{2005MNRAS.362..799M} that can be approximated by the emission of a black body with a temperature of $6.0\times10^4$ K for $h\nu<13.6$eV.
The number of stellar origin \lya photons per solar mass in star formation obtained with this method is: 
\ba
Q^{\rm stellar}_{\rm Ly\alpha}&=&4.307\int_{\nu_{ly\alpha + 2\mathring{A}}}^{\nu_{ly\alpha - 2\mathring{A}}} d\nu \frac{\nu^3}{e^{h\nu/K_b T_K}-1}\ M_{\odot}^{-1}\\ \nonumber
&=&9.92\times 10^{\rm 58}\ M_{\odot}^{-1}.
\label{eq:Qstellar_cont_lya}
\ea
We note that we are not accounting for the higher opacity at the center of the \lya line which should push the photons out of the line center before exiting the star and so we may be overestimating the stellar \lya photon emission.

Free-bound emission and free-free emission are respectively originated when free electrons scatter off ions with or without being captured.
Following the approach of \citep{2006ApJ...646..703F}, the free-free and free-bound continuum luminosity can be obtained using:
\be
L_{\nu}(M,z)=V_{sphere}(M,z)\varepsilon_{\nu}
\ee
where $V_{sphere}$ is the volume of the Str$\ddot{o}$mgren sphere which can be roughly estimated using the ratio between the number of ionizing photons emitted and the number density of recombinations in the ionized volume,
\be
V_{sphere}(M,z)=\frac{Q_{ion}SFR(M,z)(1-f_{esc})}{n_e n_p \alpha_{\beta}}.
\ee
$\varepsilon_{\nu}$ is the total volume emissivity of free-free and free-bound emission, $n_p$ is the number density of protons (ionized atoms) and $\alpha_i$ is the case A or case B recombination coefficient (see \cite{2006PhR...433..181F}).
 
The volume emissivity estimated by \citep{2003adu..book.....D} is given by:
\be
\varepsilon_{\nu}=4\pi n_e n_p \gamma_c \frac{e^{-h\nu/kT_K}}{T_K^{1/2}}J cm^{-3} s^{-1} Hz^{-1}, 
\ee
where $\gamma_c$ is the continuum emission coefficient including free-free and free-bound emission given in SI units by:
\be
\gamma_c=5.44\times 10^{-46}\left[\bar{g}_{ff}+\Sigma_{n=n\prime}^{\infty}\frac{x_n e^{x_n}}{n}g_{fb}(n)\right].
\ee
In here, $x_n=Ry/(k_B T_K n^2)$ ($k_B$ is the Boltzmann constant, n is the level to which the electron recombines to and $Ry=13.6eV$ is the Rydberg unit of energy), $\bar{g}_{ff}\approx1.1-1.2$ and $g_{fb}(n)\approx1.05-1.09$ are the thermally average Gaunt factors for free-free and free-bound emission \citep[values from]{1961ApJS....6..167K}. The initial level $n\acute{}$ is determined by the emitted photon frequency and satisfies the condition $cR_{\infty}/n\acute{} ^2 <\nu < c R_{\infty}/(n\acute{}-1)^2$ where $R_{\infty}=1.1\times10^7m^{-1}$ is the Rydberg constant. 

The continuum luminosity per frequency interval ($L_{\nu}$) is related to the \lya luminosity emitted from the galaxies by: $L_{cont}=L_{\nu}\times d\nu(4\mathring{A})=f_{Ly\alpha}Q_{Ly\alpha}E_{Ly\alpha}SFR(M,z)$, where $Q_{Ly\alpha}$ is the number of emitted lyman alpha photons per solar mass in star formation. We then obtain $Q^{\rm free-free}_{\rm Ly\alpha}=2.13\times 10^{\rm 53}$ M$_{\odot}^{-1}$ for free-free emission and $Q^{\rm free-bound}_{\rm Ly\alpha}=2.22\times 10^{\rm 55}$ M$_{\odot}^{-1}$ for free-bound emission.

During recombination there is also the probability of two photon emission and although this photons have frequencies below the lyman alpha frequency there is a small fraction of them of $Q^{\rm 2-photon}_{\rm Ly\alpha}$ that are emitted so close to the lyman alpha line that are included in the lyman alpha intrinsic width.   

The number of lyman alpha photons that can be originated due to two photon emission during recombination is given by:
\be
Q^{\rm 2-photon}_{\rm Ly\alpha}=\int_{\nu_{ly\alpha + 2\mathring{A}}}^{\nu_{ly\alpha}}\frac{2}{\nu_{Ly\alpha}}P(\nu/\nu_{Ly\alpha})d\nu,
\ee
where P(y)dy is the normalized probability that in a two photon decay one of them is the range $dy=d\nu/\nu_{ly\alpha}$ and $1-f_{ly\alpha} \approx 1/3$ is the probability of 2 photon emission during an hydrogen n=2$\rightarrow$1 transition. 
The probability of two photon decay was fitted by \cite{2006ApJ...646..703F} using Table 4 of Brown $\&$ Mathews (1970) as:
\ba
P(y)&=&1.307-2.627(y-0.5)^2+2.563(y-0.5)^4\\ \nonumber
    &-&51.69(y-0.5)^6
\ea

Finally, the different contributions to the total \lya luminosity from galaxies due to continuum emission, $L^{\rm GAL}_{\rm cont}=L^{\rm stellar}_{\rm cont}+L^{\rm free-free}_{\rm cont}+L^{\rm free-bound}_{\rm cont}+L^{\rm 2-photon}_{\rm cont}$, are given by:
\ba
L^{\rm stellar}_{\rm cont}(M,z)&=&f_{\rm Ly\alpha}Q^{\rm stellar}_{\rm Ly\alpha}E_{\rm Ly\alpha}{\rm SFR}(M,z)\\ \nonumber
&\approx& 5.12\times 10^{40}f_{\rm Ly\alpha}\frac{{\rm SFR}(M,z)}{{\rm M}_\odot\, {\rm yr}^{-1}} {\rm erg}\ s^{-1}
\ea
for stellar emission,
\ba
L^{\rm free-free}_{\rm cont}(M,z)&=&f_{\rm Ly\alpha}Q^{\rm stellar}_{\rm Ly\alpha}E_{\rm Ly\alpha}{\rm SFR}(M,z)\\ \nonumber
&\approx& 1.10\times 10^{35}f_{\rm Ly\alpha}\frac{{\rm SFR}(M,z)}{{\rm M}_\odot\, {\rm yr}^{-1}} {\rm erg}\ s^{-1}
\ea
for free-free emission,
\ba
L^{\rm free-bound}_{\rm cont}(M,z)&=&f_{\rm Ly\alpha}Q^{\rm stellar}_{\rm Ly\alpha}E_{\rm Ly\alpha}{\rm SFR}(M,z)\\ \nonumber
&\approx& 1.47\times 10^{37}f_{\rm Ly\alpha}\frac{{\rm SFR}(M,z)}{{\rm M}_\odot\, {\rm yr}^{-1}} {\rm erg}\ s^{-1}
\ea
for free-bound emission and
\ba
L^{\rm 2-photon}_{\rm cont}(M,z)&=&f_{\rm Ly\alpha}Q^{\rm stellar}_{\rm Ly\alpha}E_{\rm Ly\alpha}{\rm SFR}(M,z)\\ \nonumber
&\approx& 2.41\times 10^{38}f_{\rm Ly\alpha}\frac{{\rm SFR}(M,z)}{{\rm M}_\odot\, {\rm yr}^{-1}} {\rm erg}\ s^{-1}
\ea
for 2-photon emission.

Note that here we are only considering the part of the continuum emission from galaxies that could contribute to the same "\lya redshift". There will be a continuum emission spectrum with frequencies below the \lya line from the mechanisms above that will contribute to the same observation from lower redshifts and will generate a "foreground" to the \lya signal that needs to be removed. This should be possible due to the smoothness of this background across frequency, in the same manner as foregrounds of the 21-cm signal are removed (e.g. \citealt{2006ApJ...650..529W}).

\subsection{Modeling the relation between star formation rate and halo mass}
\label{sec:SFRvsMhalo}

Simulations of galaxy formation and observations indicate that the star formation of a halo increases strongly for small halo masses but at high halo masses ($M \gtrsim 10^{11} M_{\odot}$) it becomes almost constant \citep{2009ApJ...696..620C,2012A&A...537A..58P}.
 
In order to better estimate and constrain the SFR of a halo we used three non linear SFR versus Halo Mass parameterizations that are in good agreement with different observational constraints. In \textit{Sim1} we adjusted the SFR to reproduce a reasonable reionization history and a \lya Luminosity Function evolution compatible with different observational constraints, in \textit{Sim2} we adjusted the SFR vs halo mass relation to the parameterizations from the \cite{2011MNRAS.413..101G} galaxies catalogue (low halo masses) and the \cite{2007MNRAS.375....2D} galaxies catalogue (high halo masses). \textit{Sim2} results in an early reionization history with an optical depth to reionization compatible with the low bound of the current observational constraints. Finally \textit{Sim3} has the same halo mass dependence as \textit{Sim2} but evolves with redshift in a similar way to the \cite{2007MNRAS.375....2D} and to the \cite{2011MNRAS.413..101G} galaxy catalogues. 

We parametrized the relations between the SFR and halo mass as:     

\ba
\frac{{\rm SFR}(M,z)}{\rm M_\odot/yr}&=&\left(2.8\times10^{-28}\right) M^{a} \times \nonumber \\
&&\left(1+\frac{M}{c_1}\right)^{b}\left(1+\frac{M}{c_2}\right)^{d},
\label{SFR_param1}
\ea
where $a=2.8$, $b=-0.94$, $d=-1.7$ , $c_1=1\times10^9$ M$_\odot$ and $c_2=7\times10^{10}$ M$_\odot$ for \textit{Sim1},

\ba
\frac{{\rm SFR}(M,z)}{\rm M_\odot/yr}&=&1.6\times10^{-26}M^{a}\times \nonumber \\
&&\left(1+\frac{M}{c_1}\right)^{b}\left(1+\frac{M}{c_2}\right)^{d}\left(1+\frac{M}{c_3}\right)^{e},
\label{SFR_param2}
\ea

where $a=2.59$, $b=-0.62$, $d=0.4$, $e=-2.25$, $c_1=8\times10^8$ M$_\odot$, $c_2=7\times10^{9}$ M$_\odot$ and $c_3=1\times10^{11}$ M$_\odot$ for \textit{Sim2} and 

\ba
\frac{{\rm SFR}(M,z)}{\rm M_\odot/yr}&=&2.25\times10^{-26}\left(1+0.075\times(z-7) \right)M^{a}\times \nonumber \\
&&\left(1+\frac{M}{c_1}\right)^{b}\left(1+\frac{M}{c_2}\right)^{d}\left(1+\frac{M}{c_3}\right)^{e},
\label{SFR_param3}
\ea

where $a=2.59$, $b=-0.62$, $d=0.4$, $e=-2.25$, $c_1=8\times10^8$ M$_\odot$, $c_2=7\times10^{9}$ M$_\odot$ and $c_3=1\times10^{11}$ M$_\odot$ for \textit{Sim3}. 

\begin{figure}[htbp]
\begin{center} 
\hspace{-23pt}
\includegraphics[scale = 0.52]{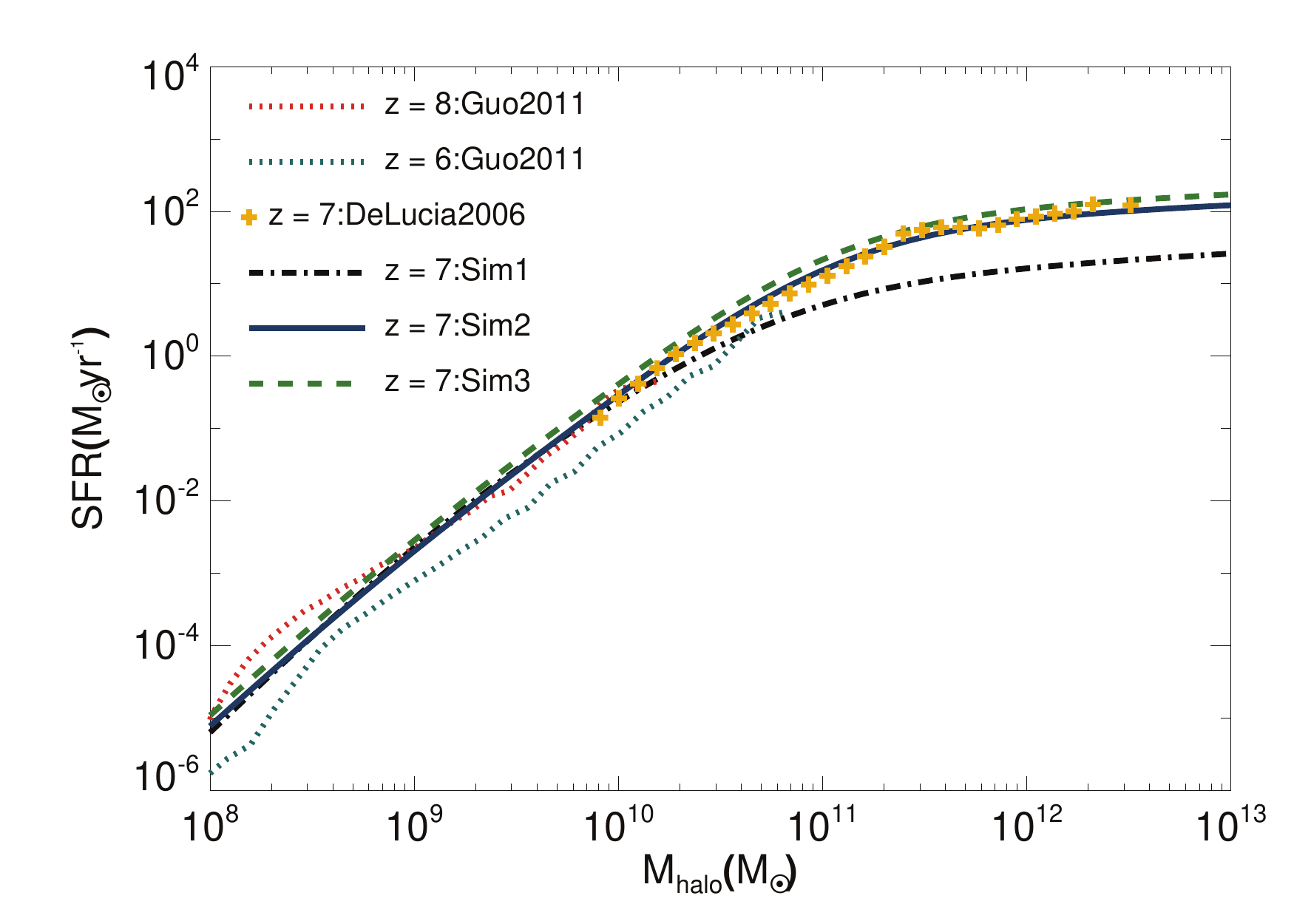} 
\caption{Star formation rate versus halo mass. The dotted lines show the relations taken from the \cite{2011MNRAS.413..101G} catalogue for low halo masses at $z=6$ (bottom dotted line) and $z=8$ (upper dotted line), the yellow crosses show the relation taken from the DeLucia catalogue for high halo masses at $z=7$. The dash-dotted, solid and dashed lines show the parameterizations used in simulations \textit{Sim1}, \textit{Sim2} and \textit{Sim3} respectively for $z=7$.}
\label{fig:SFR_halomass}
\end{center}
\end{figure}

Figure~\ref{fig:SFR_halomass} shows these relations.

\begin{figure}[htp]
\begin{center} 
\hspace{-23pt}
\includegraphics[scale = 0.52]{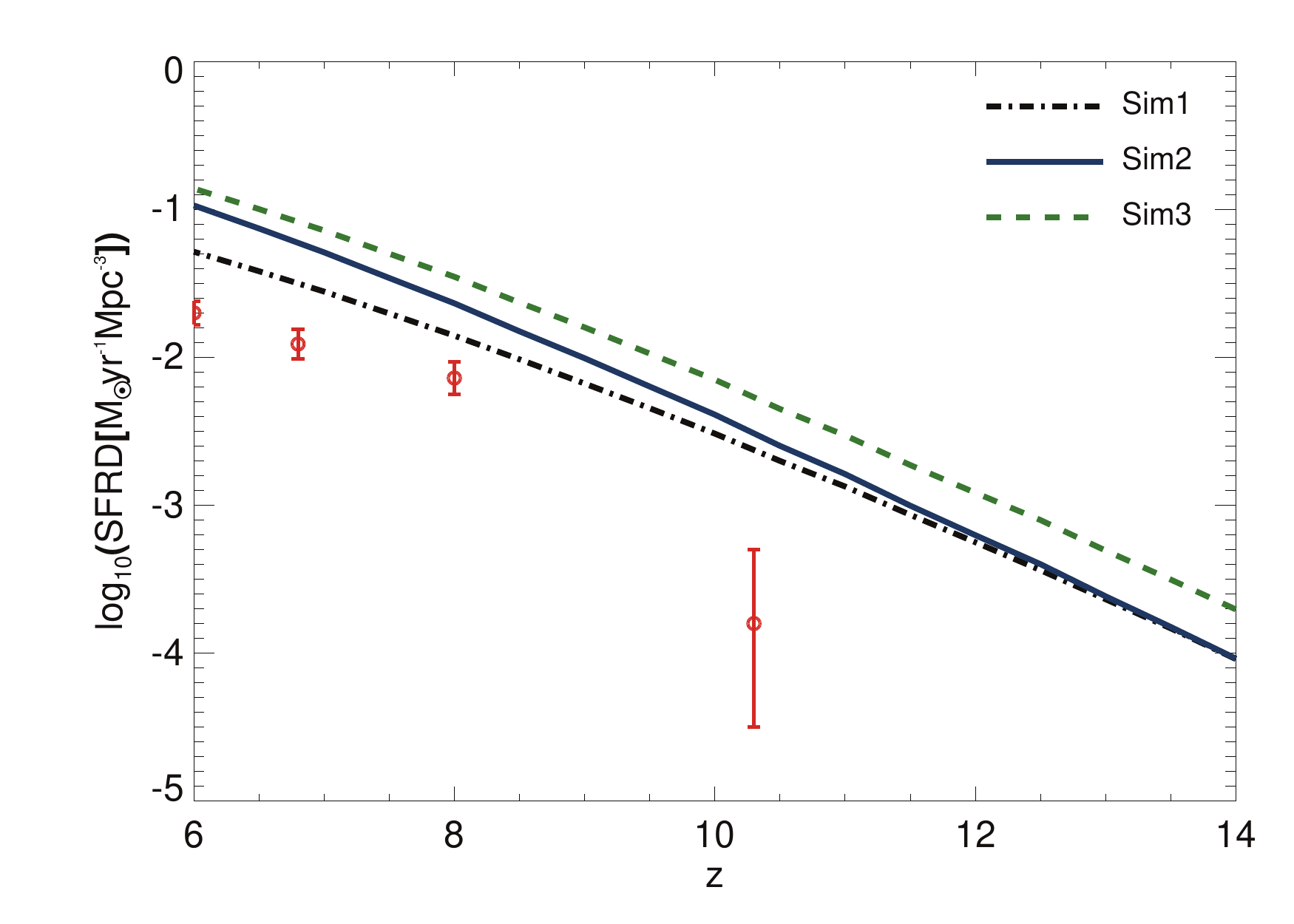} 
\caption{ Star formation rate density evolution as a function of redshift. The blue solid line, the green dashed line and the black dashed dotted line were obtained from simulations made using the SimFast21 code (for informations about the code see section \ref{sec:sim} and \citealt{2010MNRAS.406.2421S}) and the SFR vs. halo mass relations from equations \ref{SFR_param1}, \ref{SFR_param2} and \ref{SFR_param3}. The red dots are observational constraints derived from the UV luminosities corrected for dust extinction from \cite{2011arXiv1109.0994B}. Please note that these observational values correspond to high mass galaxies while our results integrate over the halo mass function starting at $\sim 10^8$ solar masses (which at redshift 7 corresponds to star formation rates of $6.41\times 10^{-5}$ $M_{\odot}$ s$^{-1}$ for \textit{Sim1}, 7.83 $\times 10^{-6}$ $M_{\odot}$ s$^{-1}$ for \textit{Sim2} and 1.1 $\times 10^{-5}$ $M_{\odot}$ s$^{-1}$ for \textit{Sim3}), so our star formation rate densities are expected to 
be higher.}
\label{fig:SFRD}
\end{center}
\end{figure}

In figure~\ref{fig:SFRD}, the strong decline in the observational SFRD from $z\approx 8$ to $z\approx 10$, imposed by the observational point at $z = 10.3$, was obtained with the observation of a single galaxy using the Hubble Deep Field 2009 two years data \citep{2011Natur.469..504B, 2012ApJ...745..110O}. It was argued in \cite{2011arXiv1105.2038B}, based on an analytical calculation, that even with such low SFRD at high redshifts it was possible to obtain an optical depth to reionization compatible with the value obtained by WMAP ($\tau=0.088\pm0.015$) \citep{2011ApJS..192...18K}. However, this derivation would imply a high escape fraction of ionizing radiation and that reionization would end at $z \approx 8$ which is hard to reconcile with the constraints from observations of quasars spectra \citep{2007ApJ...660..923M,2008arXiv0811.3918Z}. Our SFRDs are considerably higher than the current observational constraints, although the difference can be explained by a systematic underestimation of 
the SFR in observed galaxies. Moreover, current observations only probe the 
high mass end of the high redshift galaxies mass function which will underestimate the SFR density (also the obtained SFRs have very high error bars due to uncertainties in the correction due to dust extinction, the redshift and the galaxy type).
In the following sections the results shown were obtained using \textit{Sim1} unless stated otherwise.

\subsection{Total Lyman-$\alpha$ luminosity: comparison with observations}

In the previous sections we calculated the \lya luminosity as a function of the SFR for several effects.
The commonly used ``empirical'' relation between these two quantities is \citep{2011ApJ...743...65J}
\be
L_{\rm Gal}=1.1\times 10^{42}\frac{{\rm SFR}(M,z)}{{\rm M}_\odot \, {\rm yr}^{-1}} {\rm erg}\ s^{-1}
\ee
and it is based on the relation between SFR and the H$\alpha$ luminosity from \cite{1998ARA&A..36..189K} and in the line emission ratio of \lya to H$\alpha$ in case B recombinations calculated assuming a gas temperature of $10^4$ K. This empirical relation gives the \lya luminosity without dust absorption (we have labeled it {\it K98} for the remainder of the paper).

Our relation between luminosity and star formation is mass dependent (both from the escape fraction as well as due to the expression from the cooling mechanism), so in order to compare it with the result above, we calculate:
\be
A(z)=\frac{\langle L_{\rm Gal}(M,z)\rangle}{\langle {\rm SFR}(M,z) \rangle},
\label{eq:LGAL_SFR}
\ee
where the average $\langle x \rangle$ of quantity $x$ is done over the halo mass function for the mass range considered. The results are presented in table~\ref{tab:Az} for a few redshifts. 

\begin{table}
\centering                    
\begin{tabular}{l c c c c}    
\hline\hline            
A$_{\rm rec}$(z) & A$_{\rm exc}$(z) & A$_{\rm cool}$(z) & A$_{\rm cont}$(z) & A$_{\rm total}$(z) \\  
\hline                       
    $4.4\times10^{\rm 41}$ & $1.1\times10^{\rm 41}$ & $1.3\times10^{\rm 39}$ & $8.1\times10^{\rm 40}$ & $6.4\times10^{\rm 41}$ \\
    $1.2\times10^{\rm 42}$ & $3.2\times10^{\rm 41}$ & $7.8\times10^{\rm 38}$ & $6.4\times10^{\rm 40}$ & $1.6\times10^{\rm 42}$\\
    $9.3\times10^{\rm 41}$ & $2.4\times10^{\rm 41}$ & $4.9\times10^{\rm 38}$ & $4.9\times10^{\rm 40}$ & $1.2\times10^{\rm 42}$\\
    $8.5\times10^{\rm 41}$ & $2.2\times10^{\rm 41}$ & $3.3\times10^{\rm 38}$ & $3.6\times10^{\rm 40}$ & $1.1\times10^{\rm 42}$  \\
\hline                                  
\end{tabular}
\caption{Average luminosity per star formation rate (in units ${\rm erg}\ s^{-1}/{\rm M}_\odot \, {\rm yr}^{-1}$) averaged over the halo mass function for redshifts 10, 9, 8 and 7, from top to bottom.}
\label{tab:Az}     
\end{table}
Although our \lya luminosities per SFR are slightly higher, at least for low redshifts, we point out that the "empirical" relation is based on a theoretical calculation that only accounts for \lya emission due to recombinations. Moreover the observational measurements of H$\alpha$ and \lya are primarily made at low redshifts, where the absorption of \lya photons by dust in galaxies is expected to be high.
Our relation has the advantage of evolving with redshift since it accounts for the evolution of the escape fraction of ionizing photons and for the evolution of the escape fraction of \lya photons. This $z$-dependence is not present in the standard empirical relation. This redshift evolution of the UV photons escape fraction is a consequence of the increase in the number of massive galaxies with more clumpy structure as the redshift decreases. The star forming regions of massive galaxies are embedded in clumps and therefore 
it becomes more difficult for the ionizing photons to escape from such dense regions \citep{2010ApJ...710.1239R, 2011MNRAS.412..411Y}. The redshift evolution of the relation presented in equation~\ref{eq:LGAL_SFR} justifies why a theoretical calibration between \lya luminosity and the SFR of a galaxy is useful for our work. 

\begin{figure}[htbp]
\begin{center}
\hspace{-23pt}
\includegraphics[scale = 0.52]{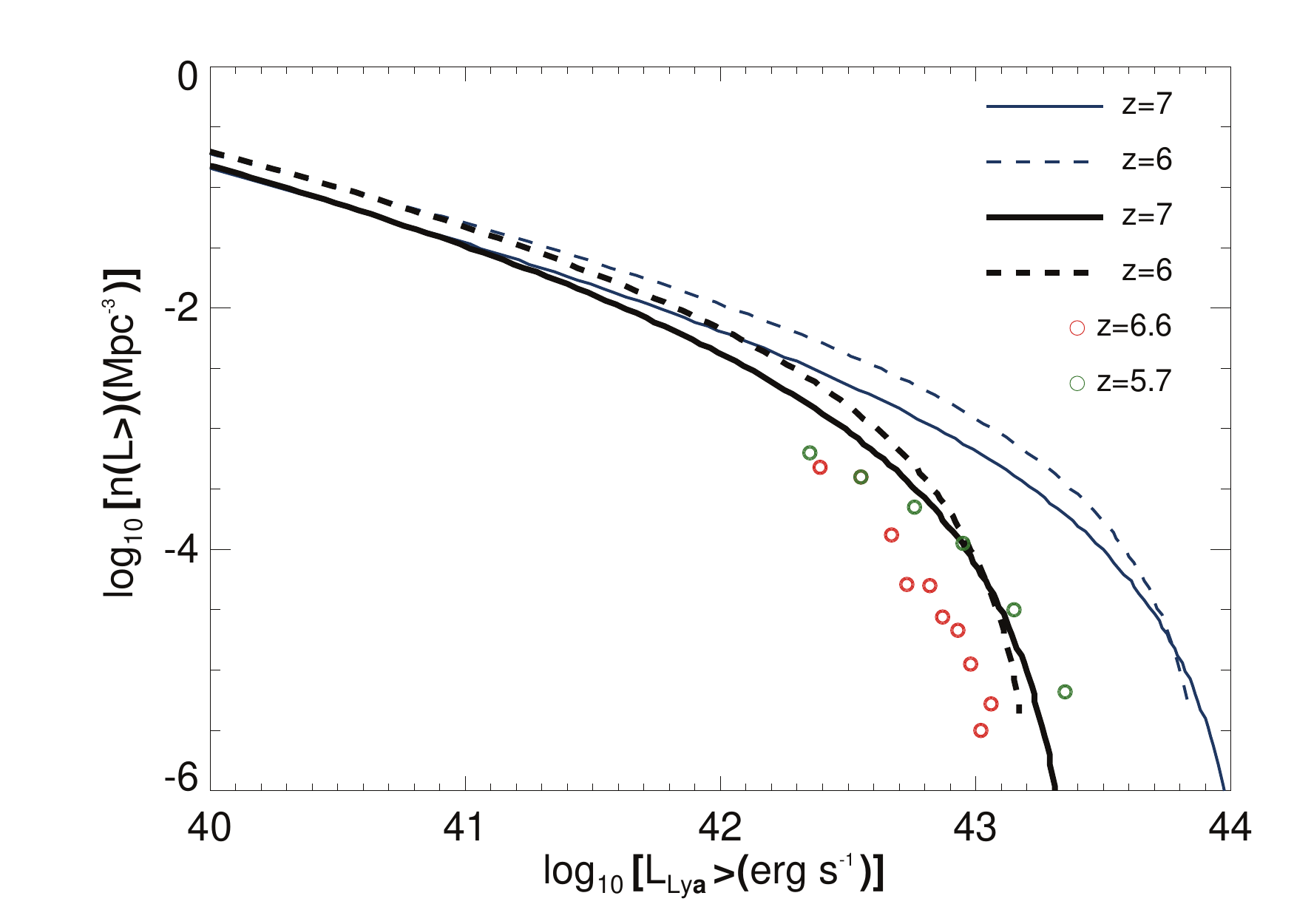} 
\caption{\lya luminosity functions obtained with our calculations are shown for redshifts $z=6$ (dashed lines) and $z=7$ (solid lines) for \textit{Sim1} (black thick lines) and \textit{Sim2} (blue thin lines). The green and red circles show the intrinsic (i.e. not affected by the IGM) \lya LF from photometric identified objects in \cite{2006PASJ...58..313S} and in \cite{2006ApJ...648....7K} for $z=5.7$ and $6.6$ respectively.}
\label{fig:LF_z6_z7}
\end{center}
\end{figure}

To check the consistency between our theoretical estimation of the \lya luminosity and the existing observations during reionization we show in figure~\ref{fig:LF_z6_z7} the luminosity function (LF) using two of the star formation rate vs. halo mass parameterizations presented in section~\ref{sec:SFRvsMhalo}.
This prediction is then compared to \lya luminosity functions of photometric identified objects in \cite{2006PASJ...58..313S} and in \cite{2006ApJ...648....7K} near the end of the reionization epoch.

Our luminosity functions were calculated assuming a minimum halo mass of $8\times10^8M_{\odot}$ which corresponds to a minimum luminosity of $3.72\times10^{36}$ erg s$^{-1}$ for \textit{Sim1}, $4.49\times10^{36}$ erg s$^{-1}$ for \textit{Sim2} and $6.22\times10^{36}$ erg s$^{-1}$ for \textit{Sim3}. 
The agreement between our LFs and observations is reasonable for \textit{Sim1} however our \textit{Sim2} overpredicts the abundance of high luminosity \lya emitters. 
This difference can be due to sample variance or a result of the high sensitivity of theoretical predictions to several parameters in our model. We point out that the luminosity range relevant for this comparison falls in a halo mass range outside the one for which the escape fraction of UV radiation we are using was estimated, so we could easily get a better fit between observations and \textit{Sim2} by reducing this escape fraction for high halo masses. This difference could also be related with the choice of halo mass function. Here we choose the Sheth-Tormen halo mass function \citep{1999MNRAS.308..119S} which has been shown to fit low-redshift simulations 
more accurately, but it is yet to be established the extent to which such a halo mass function can reproduce the halo distribution during reionization. 
Other possible explanation for this difference is the existence of a small amount of neutral gas in the IGM which would severely decrease the observed \lya luminosity from galaxies. Also, we could have decreased the high luminosity end of our luminosity functions if we had use an \lya escape fraction that decreased with halo mass such as the one used in \citep{2011MNRAS.415.3666F}.
We do not consider a model fit to the data to optimize various parameters in our model given that the current constraints on the observed \lya LFs have large overall uncertainties, especially considering variations from one survey to another.


\subsection{Lyman-$\alpha$ Average Intensity} 

In this section and the next one we will attempt to estimate the intensity and power spectrum of the \lya signal using an analytical model. In section~\ref{sec:sim} we will improve the estimation by doing the same calculation using a semi-numerical simulation.

The total intensity of \lya emission can be obtained from the combined luminosity of \lya photons associated with different mechanisms described in the previous sub-sections, such that: 
\be
\bar{I}_{\rm Gal}(z)=\int^{\rm M_{\rm max}} _{\rm M_{\rm min}}dM\frac{dn}{dM}\frac{L_{\rm Gal}(M,z)}{4\pi D_{\rm L}^{\rm 2}}y(z)D^2_A
\label{Int_Lyalpha}
\ee
where $dn/dM$ is the halo mass function \citep{1999MNRAS.308..119S}, $M$ is the halo mass, $M_{\rm max}=10^{13}M_{\rm \odot}$, $M_{\rm min}=M_{\rm OB}$, $D_{\rm L}$ is the proper luminosity distance and $D_{\rm A}$ the comoving angular diameter distance. Finally, $y(z)=d\chi/d\nu=\lambda_{\rm Ly\alpha}(1+z)^{\rm 2}/H(z)$, where $\chi$ is the comoving distance, $\nu$ is the observed frequency and $\lambda_{\rm Ly\alpha}=2.46\times10^{-15}$ m is the rest-frame wavelength of the \lya line. 

\begin{figure}[htbp]
\begin{center}
\hspace{-19pt}
\includegraphics[scale = 0.51]{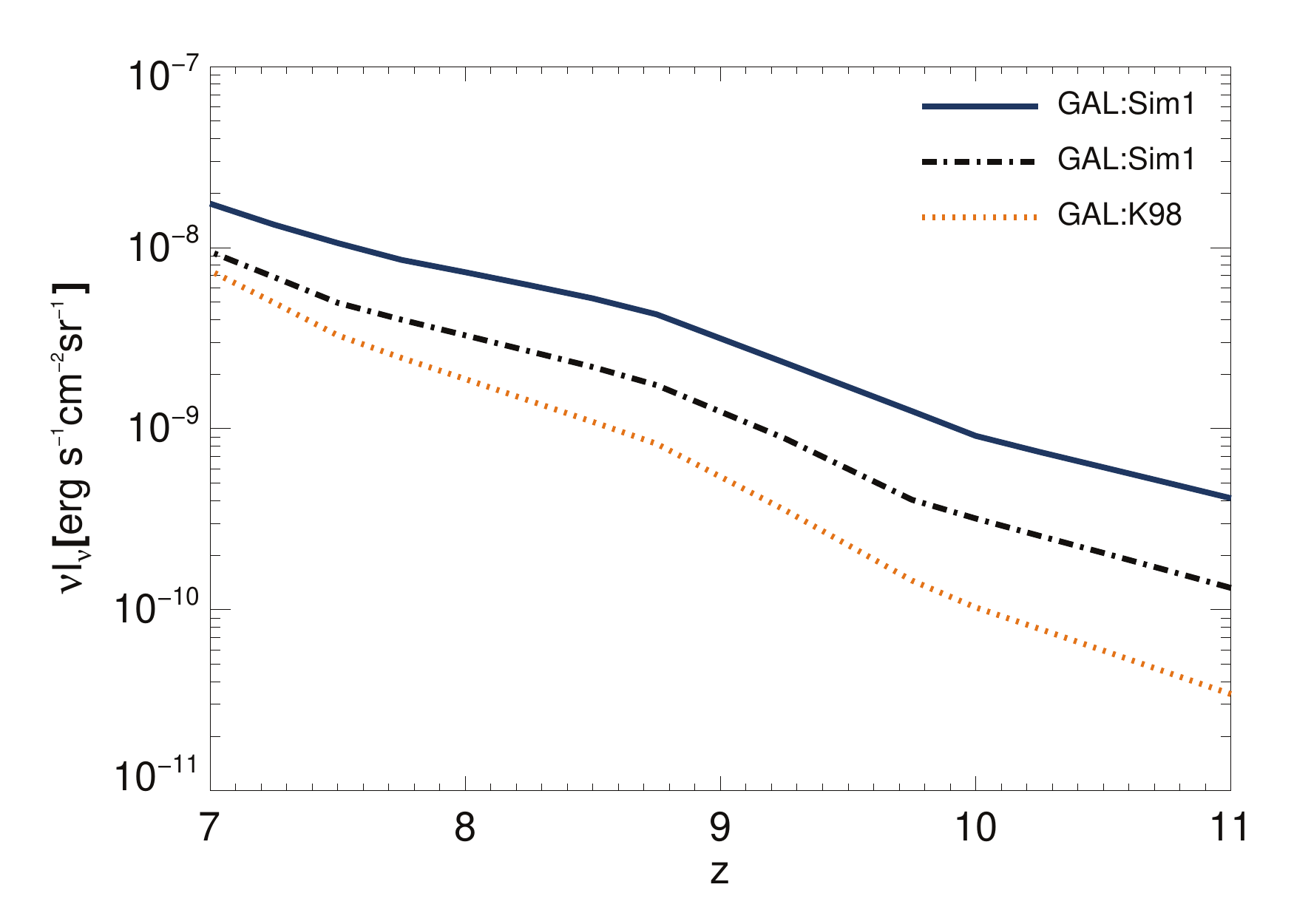} 
\caption{\lya Intensity from galaxies in erg s$^{-1}$\,cm$^{-2}$\,sr$^{-1}$ as a function of redshift. The black dashed dotted line and the blue solid line were obtained using our theoretical calculation of the \lya luminosity and the SFR halo mass relation from \textit{Sim1} and \textit{Sim2} respectively. The orange dotted line uses the \lya luminosity SFR relation based on the relation between SFR and the H$\alpha$ luminosity from \cite{1998ARA&A..36..189K} and the line emission ratio of \lya to H$\alpha$ in case B recombinations calculated assuming a gas temperature of $10000$K (labeled as the K98 relation). The K98 line is not corrected for dust absorption.}
\label{fig:Int_GAL}
\end{center}
\end{figure}

The evolution of the \lya intensity predicted by this calculation is shown in figure~\ref{fig:Int_GAL} together with the scaling expected under the ''empirical'' relation from  \cite{1998ARA&A..36..189K}
combined with an assumption related to the gas temperature.
The intensities of \lya emission from different sources are presented in table~\ref{table_GAL} for several redshifts.

\begin{table}[h!]
\centering                         
\begin{tabular}{l  c c c}        
\hline\hline                 
Source of emission in  & $\nu$I$_\nu$(z=7)  & $\nu$I$_\nu$(z=8) & $\nu$I$_\nu$(z=10) \\
$[{\rm erg\ s^{-1} cm}^{\rm -2}{\rm sr}^{\rm -1}]$ \\   
\hline                      
   Recombinations & $7.3\times10^{-9}$ & $2.5\times10^{-9}$ & $2.3\times10^{-10}$\\ 
   Excitations    & $1.9\times10^{-9}$ & $6.5\times10^{-10}$ & $6.0\times10^{-11}$\\
   Cooling        & $2.8\times10^{-12}$ & $1.5\times10^{-12}$ & $4.7\times10^{-13}$\\
   Continuum      & $3.1\times10^{-10}$ & $3.5\times10^{-10}$ & $3.0\times10^{-11}$\\
   Total          & $9.5\times10^{-9}$ & $3.5\times10^{-9}$ & $3.2\times10^{-10}$\\
\hline                                  
\end{tabular}
\caption{Surface brightness (in observed frequency times intensity) of \lya emission from the different sources in galaxies at $z\approx 7$, $z\approx 8$  and $z\approx 10$ for \textit{Sim1}.}
\label{table_GAL}     
\end{table}

These intensities can be extrapolated to other SFRDs, assuming that the only change is in the amplitude of the SFR halo mass relations presented in figure \ref{fig:SFR_halomass} by using the coefficients in table \ref{table_Int_SFRD}. 

\begin{table}[h!]
\centering                         
\begin{tabular}{l  c c c}        
\hline\hline                 
Redshift & A$_{SFRD}$(\textit{Sim1}) & A$_{SFRD}$(\textit{Sim2}) \\
\hline                      
   10 & $1.05\times10^{-03}$ & $2.20\times10^{-03}$\\ 
   9  & $2.17\times10^{-03}$ & $3.75\times10^{-03}$\\
   8  & $2.34\times10^{-03}$ & $3.15\times10^{-03}$\\
   7  & $3.42\times10^{-03}$ & $3.42\times10^{-03}$\\
\hline                                  
\end{tabular}
\caption{Average \lya Intensity from galaxies per SFRD (A$_{SFRD}$) in units ${\rm erg\ s^{-1} cm}^{\rm -2}{\rm sr}^{\rm -1}/ M_{\odot} yr^{-1}$, calculated using the star formation rate halo mass relation from simulations \textit{Sim1} and \textit{Sim2}.}
\label{table_Int_SFRD}     
\end{table}

The intensities from emission at $z\approx 7$, $8$ and $10$ are $9.5 \times 10^{-9}$, $3.5 \times 10^{-9}$ and $3.2 \times 10^{-10}$ erg s$^{-1}$\,cm$^{-2}$\,sr$^{-1}$, respectively. Such an intensity is substantially smaller than the background intensity of integrated emission from all galaxies (around $1\times10^{-5}$ erg s$^{-1}$\,cm$^{-2}$\,sr$^{-1}$ \citep{2000MNRAS.312L...9M}, or from the total emission of galaxies during reionization, estimated to be at most $1\times10^{-6}$ erg s$^{-1}$cm$^{-2}$\,sr$^{-1}$ (Yan et al. 2012).

\subsection{Lyman-$\alpha$ intensity power spectrum}
\label{sec:power}

The \lya emission from galaxies will naturally trace the underlying cosmic matter density field so we can write the \lya line intensity fluctuations due to galaxy clustering as
\be
\delta I_{\rm GAL}=b_{\rm Ly\alpha}\bar{I}_{\rm GAL}\delta({\bf x}),
\label{eq:d_Iv}
\ee
where $\bar{I}_{\rm GAL}$ is the mean intensity of the \lya emission line, $\delta({\bf x})$ is the matter over-density at the location ${\bf x}$, and $b_{\rm Ly\alpha}$ is the average galaxy bias weighted by the \lya luminosity (see e.g. \citealt{2011ApJ...728L..46G}).

\begin{figure}[t!]
\begin{center}
\hspace{-8pt}
\includegraphics[scale = 0.5]{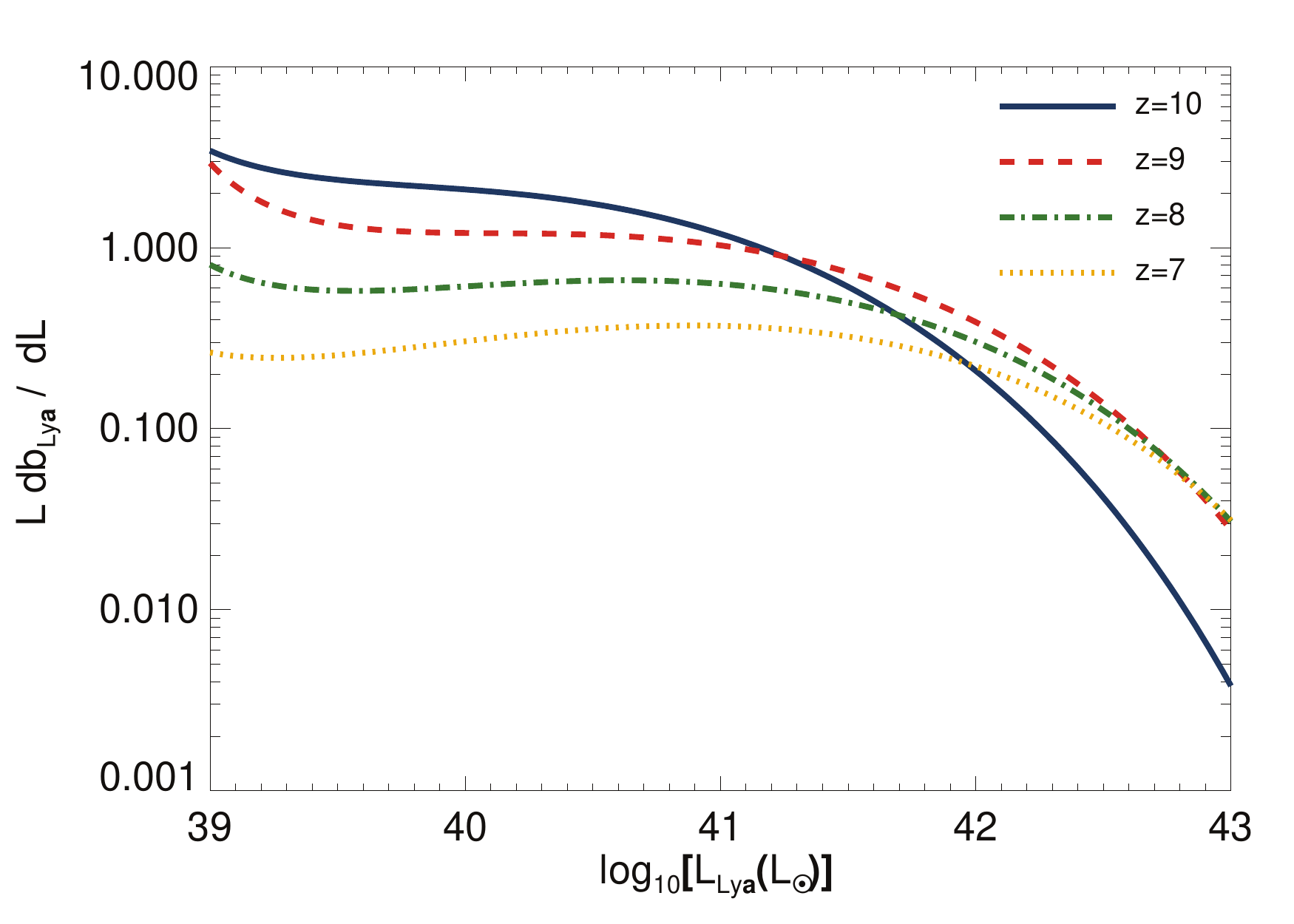} 
\caption{Bias between dark matter fluctuations and \lya surface brightness $(\nu I)$ from galaxies as a function of the galaxy Lyman alpha luminosity at redshifts $7$, $8$ ,$9$ and $10$.}
\label{fig:Bias_Lya}
\end{center}
\end{figure}

Using one of the relations between the SFR and halo mass from section \ref{sec:SFRvsMhalo} we can calculate the luminosity and obtain the Lyman alpha bias following \citet{2010JCAP...11..016V}:
\be
b_{\rm Ly\alpha}(z)=\frac{\int^{M_{\rm max}}_{M_{\rm min}} dM \frac{dn}{dM} L_{\rm GAL}\  b(z,M)}{\int^{M_{\rm max}}_{M_{\rm min}} dM \frac{dn}{dM} L_{\rm GAL}},
\label{eq:4b}
\ee 
where $b(z,M)$ is the halo bias and $dn/dM$ the halo mass function \citep{1999MNRAS.308..119S}. We take $M_{\rm min}=10^8 M_{\odot}/h$ and $M_{\rm max}=10^{13} M_{\odot}/h$. 
The bias between dark matter fluctuation and the \lya luminosity, as can be seen in figure~\ref{fig:Bias_Lya}, is dominated by the galaxies with low \lya luminosity independently of the redshift.

\begin{figure}[htbp]

\begin{center}
\hspace{-16pt}
\includegraphics[scale = 0.51]{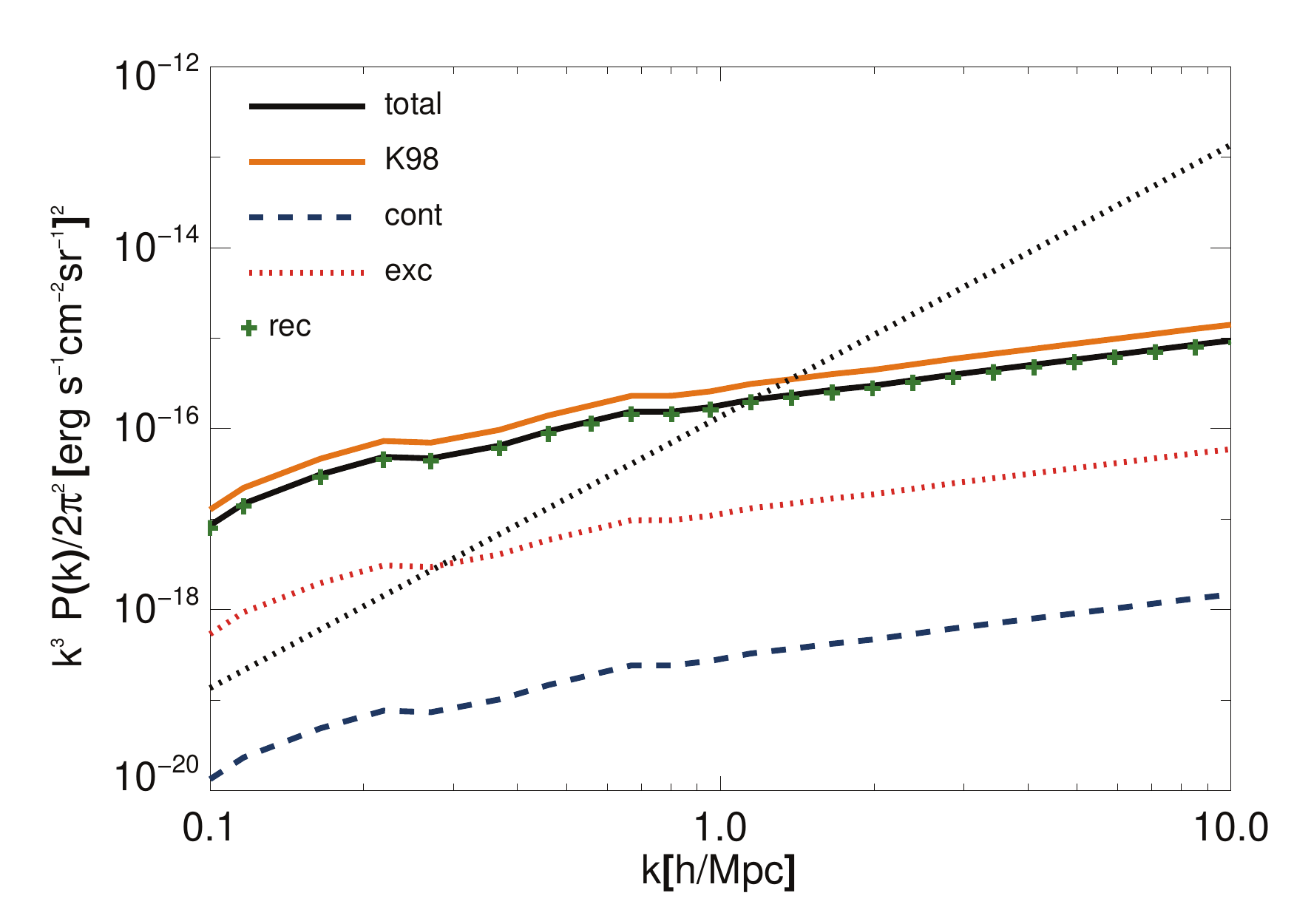}

\hspace{-16pt}
\includegraphics[scale = 0.51]{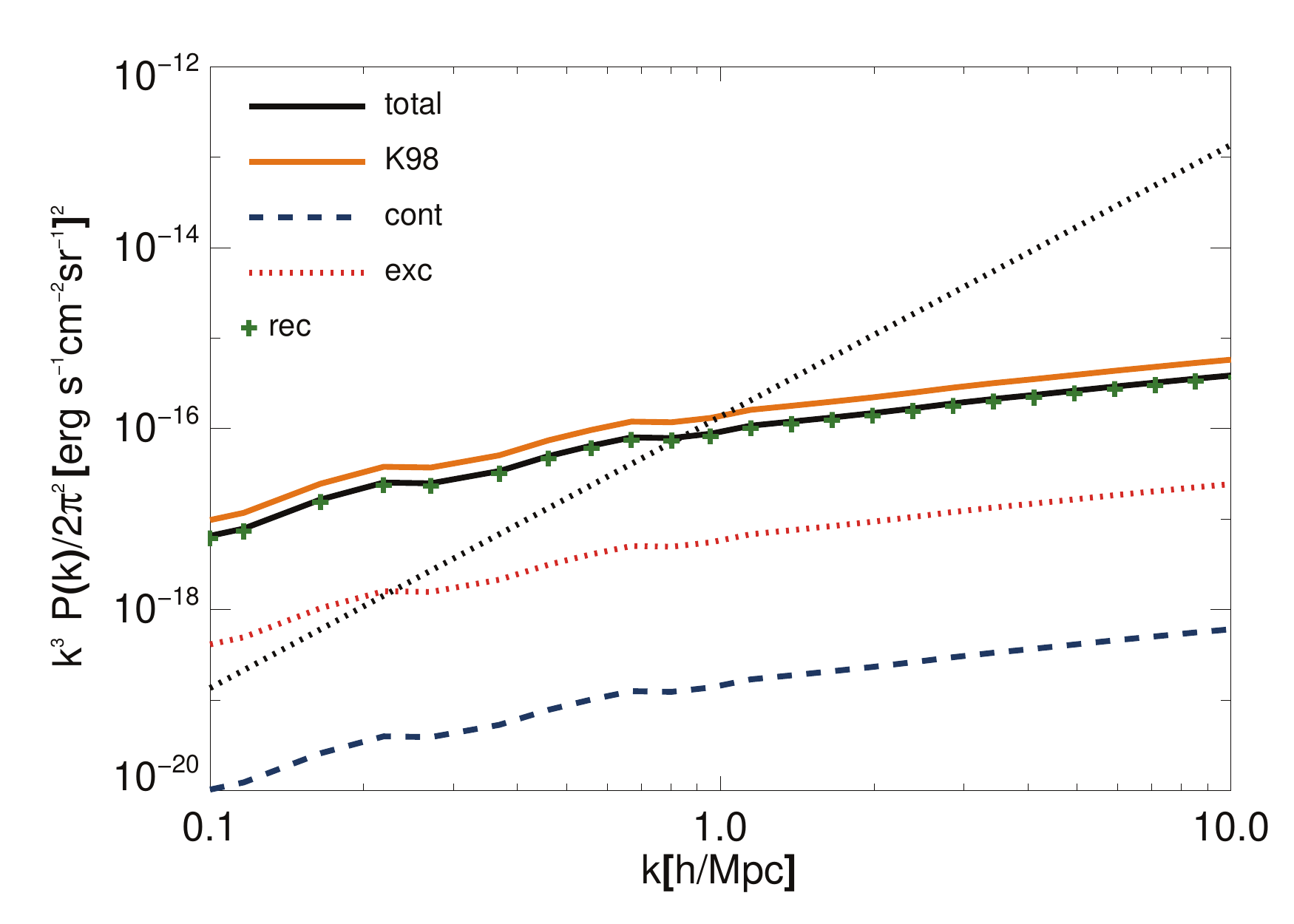}

\caption{Clustering power spectrum of the \lya surface brightness $(\nu I)$ from Galaxies at redshifts 7 to 10 (from top to bottom), from several sources: collisions and excitations, recombinations and continuum emission with frequencies inside the \lya width. The power spectra from cooling emission is not shown since it is several orders of magnitude smaller than the contributions from the other sources.  Also shown are the total power spectra (clustering (solid black line) and shot noise power spectra (dotted black line)) of the total contribution for \lya emission in galaxies predicted by our theoretical calculation and total \lya clustering power spectra predicted using the $K98$ relation (orange solid line).}
\label{fig:ps_Lya_gal_bias}
\end{center}
\end{figure}

We can then obtain the clustering power spectrum of \lya emission as
\be
P_{\rm GAL}^{\rm clus}(z,k) = b_{\rm Ly\alpha}^2 \bar{I}_{\rm GAL}^2 P_{\delta \delta}(z,k),
\ee
where $P_{\delta \delta}(z,k)$ is the matter power spectrum.
The shot-noise power spectrum, due to discretization of the galaxies, is also considered here. It can be written as \citep{2011ApJ...728L..46G}
\be
P^{\rm shot}_{\rm Ly\alpha}(z) = \int_{\rm M_{\rm max}}^{\rm M_{\rm min}} dM \frac{dn}{dM} 
\bigg(\frac{L_{\rm GAL}}{4\pi D_{\rm L}^2}y(z)D_{\rm A}^2\bigg)^2.
\ee
The resulting power spectrum of \lya emission from galaxies is presented in figure~\ref{fig:ps_Lya_gal_bias}. At all scales presented the \lya intensity and fluctuations are dominated by the recombination emission from galaxies.

\section{Lyman-$\alpha$ emission from the IGM} 
\label{sec:IGM}

The \lya emission from the IGM is mostly originated in recombinations and collisions powered by the ionizing background. These processes are similar to the ones described inside the galaxies, although, since the physical conditions of the gas in the IGM are different from those in the ISM, the intensity of \lya emission can no longer be connected to the ionizing photon intensity using the previous relations. The biggest challenge in doing these calculations is to connect the IGM ionizations and heating of the gas to the emission of ionizing radiation and the star formation rate assumed in the previous sections. Moreover, in the IGM, we also have to take into account the contribution of continuum radiation from stars between the \lya and the Lyman limit which redshifts into the \lya line. 

In a schematic view, we have to take into account the following processes,
\begin{enumerate}
\item The amount of energy in UV photons that escapes the galaxy
\item This energy will then be distributed in the IGM into:
\begin{enumerate}
\item ionizations
\item direct excitations (followed by emission, partially into the \lya line)
\item heating of the gas
\end{enumerate}
\item Taking into account the state of the IGM in terms of temperature and ionization, we can then further determine how much it will radiate through the \lya line from:
\begin{enumerate}
\item Recombinations
\item Radiative cooling (usually through excitations followed by decay in several lines including \lya)
\end{enumerate}
\item The amount of \lyn photons that escape the galaxy, re-scattering in the IGM into \lya photons
\end{enumerate}
The proper calculation of all these processes will require simulations which we will address in section~\ref{sec:sim}. In the following sub-sections we review the contributions through analytical calculations in order to get a better understanding of the dominating effects.

\subsection{Lyman-$\alpha$ emission from hydrogen recombinations}
\label{sec:3.1}

The UV radiation that escapes the interstellar medium into the intergalactic medium ionizes low density clouds of neutral gas. Part of the gas in these clouds then recombines giving rise to \lya emission. 
The radiation emitted in the IGM is often referred to as fluorescence \citep{2004MNRAS.349.1137S}. 
The comoving number density of recombinations per second in a given region, $\dot{n}_{rec}$, is given by:
\be
\dot{n}_{rec}(z)=\alpha n_e(z) n_{\rm HII}(z)
\label{eq:nrec_s}
\ee
where $\alpha$ changes between the case A and the case B recombinations coefficient, $n_{\rm HII}=x_i \frac{n_b(1-Y_p)}{1-3/4Y_p}$ is the ionized hydrogen comoving number density ($x_i$ is the ionization fraction, $n_b$ the baryonic comoving number density). The free electron density can be approximated by $n_e=x_i n_b$.
 
The recombination coefficients are a function of the IGM temperature, $T_K$. The case A comoving recombination coefficient is appropriate for the highly ionized low redshift Universe \cite{2006PhR...433..181F},
\be
\alpha_A\approx 4.2 \times 10^{-13}(T_K/10^4{\rm K})^{-0.7}(1+z)^3\ cm^3 s^{-1}
\ee
while the case  B comoving recombination coefficient is appropriate for the high redshift Universe.
\be
\alpha_B\approx 2.6 \times 10^{-13}(T_K/10^4{\rm K})^{-0.7}(1+z)^3\ cm^3 s^{-1}.
\ee
The use of a larger recombination coefficient when the process of hydrogen recombination is close to its end accounts for the fact that at this time, ionizations (and hence recombinations) take place in dense, partially neutral gas (Lyman-limit systems) and the photons produced after recombinations are consumed inside this systems so they do not help ionizing the IGM \citep[see: eg.][]{2006PhR...433..181F}.

The fraction of \lya photons emitted per hydrogen recombination, $f_{rec}$, is temperature dependent so we used the parameterization for $f_{rec}$ made by \cite{2008ApJ...672...48C} using a combination of fits tabulated by \cite{1964MNRAS.127..145P} and \cite{1988ApJS...66..125M} for $T_K>10^3$ and $T_K<10^3$ respectively:
\be
f_{\rm rec}=0.686-0.106\log_{10}(T_K/10^4{\rm K})-0.009(T_K/10^4{\rm K})^{-0.4}.
\ee
The luminosity density (per comoving volume) in \lya from hydrogen recombinations in the IGM, $\ell^{\rm IGM}_{\rm rec}$, is then given by
\be
{\rm \ell}^{\rm IGM}_{\rm rec}(z)= f_{\rm rec} \dot{n}_{\rm rec}  E_{\rm Ly\alpha}.
\ee

\subsection{Lyman-$\alpha$ emission from excitations in the IGM}
\label{sec:3.2}

The UV radiation that escapes the galaxies without producing ionization ends up ionizing and exciting the neutral hydrogen in the IGM and heating the gas around the galaxies. The high energetic electron released after the first ionization spends its energy in collisions/excitations, ionizations and heating the IGM gas until it thermalizes \citep{1985ApJ...298..268S}. We estimated the contribution of the direct collisions/excitations to the \lya photon budget and concluded that it is negligible.

The \lya luminosity density due to the collisional emission (radiative cooling in the IGM), $\ell^{\rm IGM}_{\rm exc}$, is given by:   
\be
{\rm \ell}^{\rm IGM}_{\rm exc}(z)=  n_e n_{\rm HI}q_{\rm Ly\alpha}  E_{\rm Ly\alpha},
\ee 
where $n_{\rm HI}=n_b (1-x_i)\frac{(1-Y_p)}{1-3/4Y_p}$ is the neutral hydrogen density, $x_i$ is the IGM ionized fraction and $q_{\rm Ly\alpha}$ is the effective collisional excitation coefficient for \lya emission which we calculated in the same way as \cite{2008ApJ...672...48C}, but using different values for the gas temperature and IGM ionization fraction.

Considering excitation processes up to the level $n=3$ that could eventually produce \lya emission, the effective collisional excitation coefficient is given by:
\be
q_{\rm Ly\alpha}=q_{1,2p}+q_{1,2s}+q_{1,3p}.
\ee
The collisional excitation coefficient for the transition from the ground level ($1$) to the level ($nl$) is given by
\be
q_{1,nl}=\frac{8.629\times10^{-6}}{T_K^{1/2}}\frac{\Omega(1,nl)}{\omega_1}e^{E_{1,n}/k_B T_K} {\rm cm}^{3}{\rm s}^{-1},
\ee 
where $\Omega(1,nl)$ is the temperature dependent effective collision strength, $\omega_1$ is the statistical weight of the ground state, $E_{1,n}$ is the energy difference between the ground and the $nl$ level and $k_B$ is the Boltzmann constant.

\subsection{Scattering of Lyman-n photons emitted from galaxies}
\label{sec:3.3}
Continuum emission of photons, by stars, from \lya to the Lyman-limit travels until it reaches one of the \lyn lines where it gets scattered by neutral hydrogen. Most of this scattering will have as end result the production of \lya photons which eventually redshift out of the line.
Since a considerable fraction of this photons only reach a given \lyn frequency in the IGM this \lya emission is formed as a flux that decays with $r^2$ around the star that emitted the continuum photons so it appears diluted in frequency in line observations of point sources \citep{2008ApJ...684...18C}. This continuum photons are much less likely to be absorbed by the dust in the ISM than photons originated in recombinations.
 
In intensity mapping the frequency band observed is much larger than in line observations so in principle all the continuum \lya photons can be detected. 
Using the Spectral Energy Distribution (SED) made with the code from \cite{2005MNRAS.362..799M} we estimated that the number of photons emitted by stars between the \lya plus the lyman alpha equivalent width and the Lyman-limit is equivalent to $Q^{\rm IGM}_{\rm Lyn}=9.31\times 10^{60} M_{\odot}^{-1}s^{-1}$. The higher frequency photons are absorbed by hydrogen atoms as they reach the Lyman beta frequency, reemitted and suffer multiple scattering until they reach the \lya line.  The fraction of the continuum photons emitted close to the \lya line have already redshifted to lower frequencies before reaching the IGM so they will not be scattered by the neutral hydrogen in the IGM and will not contribute to the radiative coupling of the 21 cm signal (they are already included in the calculation of the \lya emission from galaxies).

The intensity of this emission was calculated with a stellar emissivity that evolves with frequency as $\nu^{-\alpha}$ with $\alpha =0.86$ and normalized to $Q^{\rm IGM}_{\rm Lyn}$. 
The \lya luminosity density originated from continuum stellar radiation and emitted in the IGM, $\ell^{\rm IGM}_{\rm cont}$, is then approximately given by:
\be
{\rm \ell^{\rm IGM}_{\rm cont}}(z)\approx Q^{\rm IGM}_{Ly\alpha}E_{\rm Ly\alpha}{\rm SFRD}(z),
\ee
where the SFRD is in units of $M_{\odot}$ per second. Note that in section \ref{sec:sim}, this calculation is done through a full simulation.

\subsection{Lyman-$\alpha$ Intensity} 
\label{sec:3.4} 
We calculated the intensities for the several \lya sources in the IGM from their luminosity densities using:
\be
\bar{I}_{\rm IGM}(z)=\frac{{\rm \ell^{\rm IGM}}(z)}{4\pi D_L^2}y(z)D^2_A.
\label{Int_Lyalpha_IGM}
\ee

\begin{figure}[htbp]
\begin{center}
\hspace{-18pt}
\includegraphics[scale = 0.51]{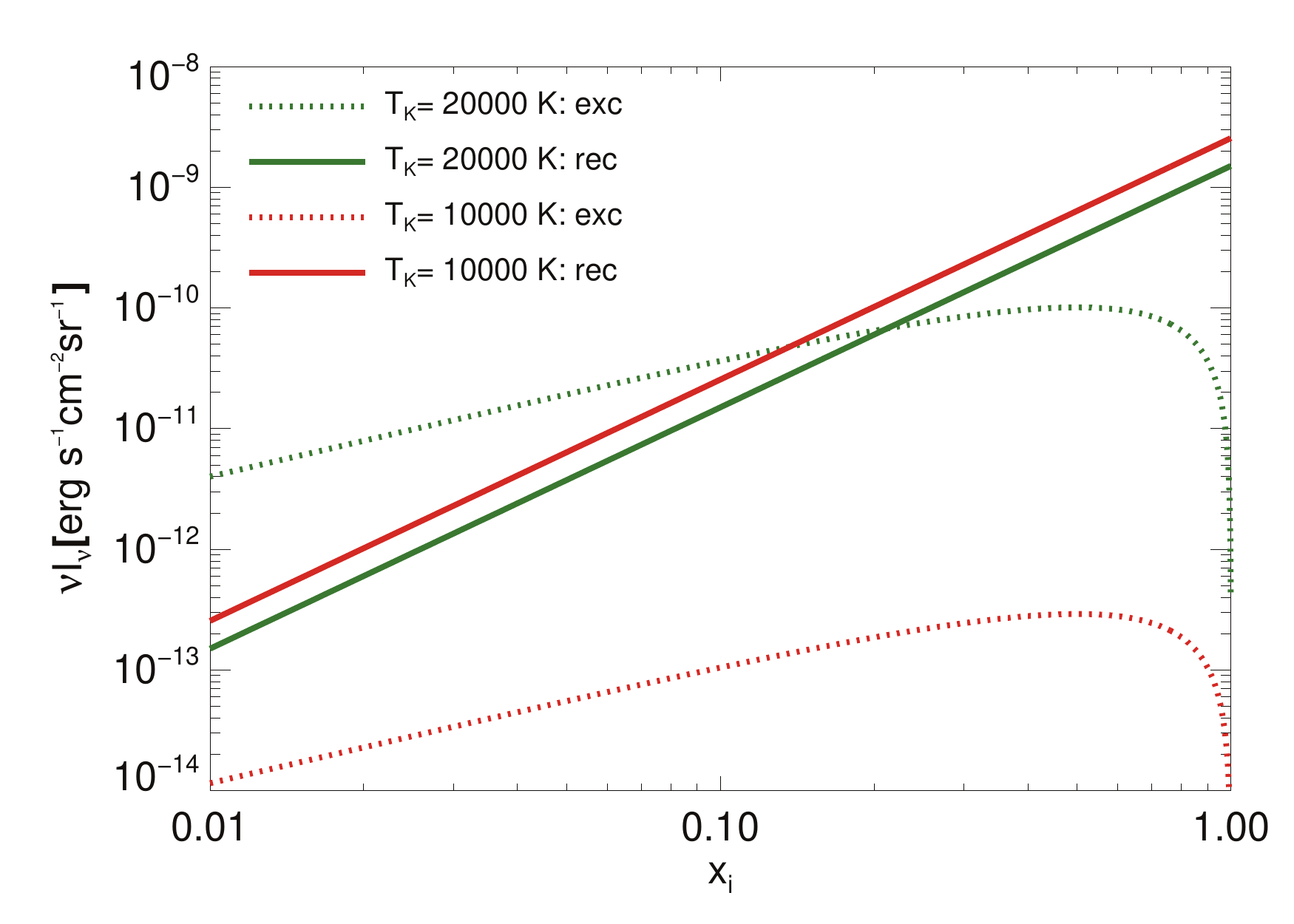} 
\caption{Intensity of \lya emission at redshift 7 due to recombinations and excitations in the IGM as a function of the hydrogen ionized fraction. The green and red lines assume a constant gas temperature of 20000 K and 10000 K respectively.} 
\label{I_tk_xi1}
\end{center}
\end{figure}

The luminosity and so the intensity of \lya emission in the IGM depends on local values of the hydrogen ionized fraction, the gas temperature and the gas density. These parameters are correlated with each other and so theoretical calculations of the average intensity made with the average of this parameters may be misleading. Since this emission is dominated by overdense regions a clumping factor of a few units is usually assumed in theoretical calculations. However we decided to estimate this intensity without using a clumping factor since its effect can be extrapolated from the intensity without clumping. The intensity of \lya emission due to recombinations or collisions in the IGM is shown in figure~\ref{I_tk_xi1} as a function of the hydrogen ionized fraction for different values of the gas temperature. 

Even for a fixed average IGM ionized fraction, the intensity of \lya emission is the result of emission from several regions and so all the values shown in figure~\ref{I_tk_xi1} are relevant.
As can be seen in figure~\ref{I_tk_xi1}, the intensity of Lyman alpha due to recombinations and collisions in the IGM is very sensitive to the gas temperature and to the fluctuations in the IGM ionized fraction. Numerical simulations predict that the temperatures in the hydrogen gas in the IGM can vary in the range 5000 K to 20000 K \citep{2001ApJ...552..473D,2011ApJ...731....6S}.

\begin{figure}[htbp]
\begin{center}
\hspace{-18pt}
\includegraphics[scale = 0.51]{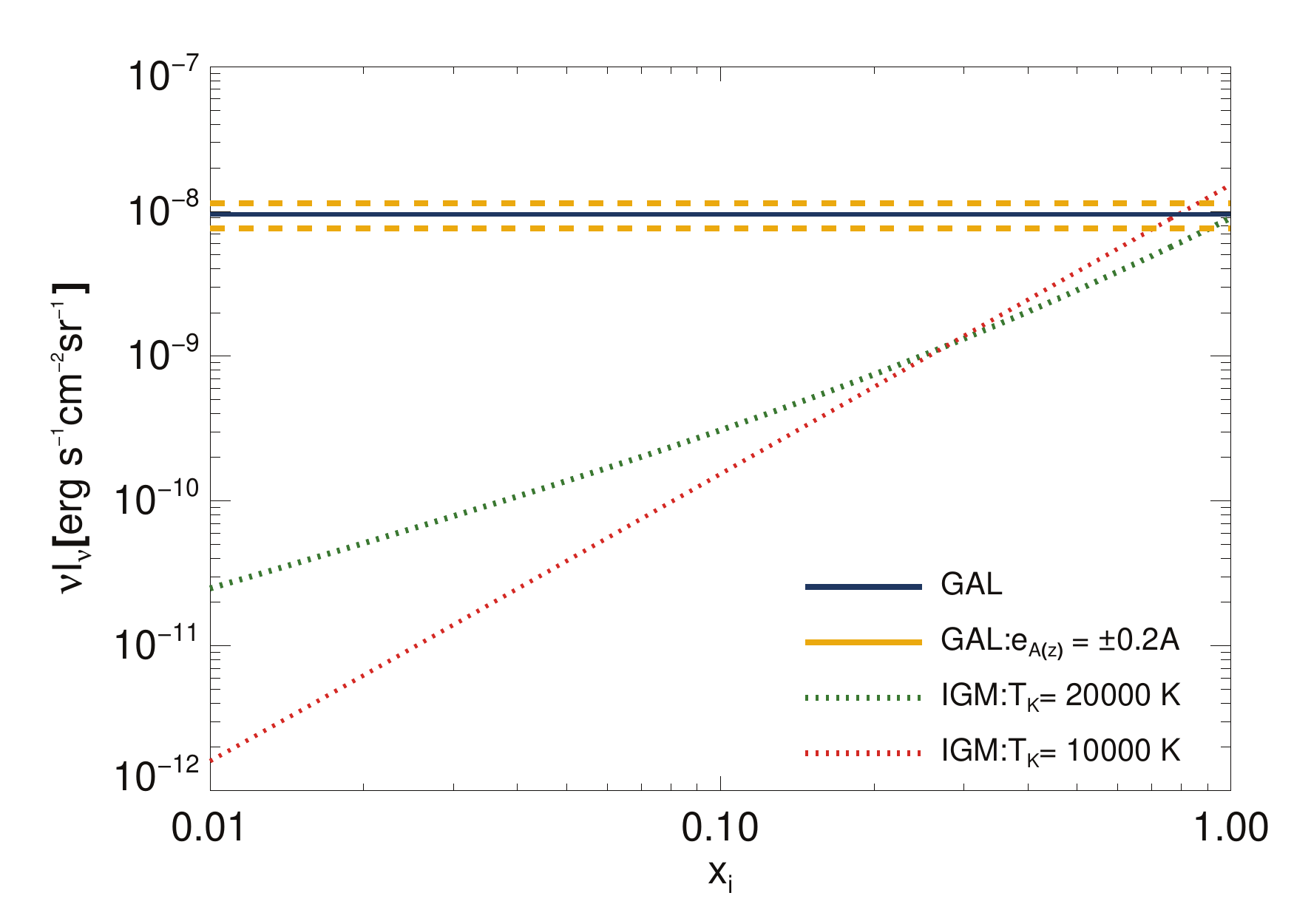} 
\caption{Intensity of \lya emission at redshift seven from the IGM and from Galaxies as a function of the hydrogen ionized fraction and including all contributions. The green and red lines are the intensity of \lya emission in the IGM assuming a constant gas temperature of $20000$ K and $10000$ K respectively. The blue solid line is the intensity of \lya emission from Galaxies as calculated in the previous section. The yellow dashed lines show the intensity in galaxies assuming an error in A(z) of 20\% due to the uncertainty in the ionizing photons escape fraction and due to the uncertainty in the 
emissivity 
of ionizing photons. The intensities in the IGM were calculated assuming a clumping factor of $6$ compatible with current conservative estimates.\citep{2010ASPC..432..230P}.} 
\label{I_tk_xi2}
\end{center}
\end{figure}

The theoretical intensities of \lya emission in galaxies and in the IGM shown in figure~\ref{I_tk_xi2} indicate that unless the IGM clumping factor is very high, or the \lya photon escape fraction is very low, the Lyman alpha intensity from the IGM at $z=7$ is lower than the emission from galaxies. 

At higher redshifts the SFRD will decrease causing the \lya intensities from galaxies and from the IGM to decrease. The escape fraction of UV photons from galaxies increases as the redshift increases which will contribute negatively to the intensity of emission in galaxies and positively to the intensity of emission in the IGM. At high redshifts the IGM ionized fraction is small which contributes to a strong decrease in the intensity of emission from the IGM compared to the intensity at $z=7$.

\section{Lyman-$\alpha$ intensity and power spectrum using numerical simulations}
\label{sec:sim}

The intensity of Lyman alpha emission in the IGM at a given time and a given region is proportional to the ionized fraction, the gas temperature and the matter density in that region. Since these three quantities are correlated, the use of average values in the \lya intensity calculation highly underestimates the emission in the more overdense regions. Also the evolution of the average of the IGM ionized fraction is poorly known during the Epoch of Reionization (EOR).

Some of these problems can be resolved using a computational code able to produce simulations of the IGM ionized fraction, the gas temperature and the matter density in a volume high enough to properly represent our Universe. The use of simulations has an additional advantage of allowing the calculation of the 3 dimensional power spectra of Lyman alpha emission in the IGM without the need for assuming a bias relation with the underlying dark matter distribution.

In this section we will estimate the inhomogeneous \lya intensity from Galaxies and the IGM using a modified version of the SimFast21 code \citep{2010MNRAS.406.2421S}. 
Given a set of astrophysical and cosmological parameters, this code is able to consistently produce 3 dimensional simulations of the dark matter density field, the ionization field, the SFRD, the scattering of \lyn photons in the IGM, the X-ray heating of the IGM and even 21 cm spin and brightness temperature fluctuations for the several redshifts of the EoR. 

A proper calculation of all the heating and cooling mechanisms would add a high level of complexity to this calculation and would require a small redshift step in the IGM fraction calculation so we assumed a constant temperature in ionized regions of 10000K. Moroever, the results from our calculations can be easily extrapolated to account for a higher temperature.
For example for a temperature of 20000K the number of recombinations in the IGM would decrease by a factor of 1.7 and the number of collisions would increase more than two orders in magnitude. Assuming that the clumping of the IGM is not very high, and so \lya recombination emission dominates over collisional emission during most of the EOR, than this higher temperature would cause a small decrease in the intensity of emission in the IGM and the Reionization period would be less extended than what we predict in section \ref{sec:SimFast_xi_Tk}. 
We made a few modifications to the SimFast21 code in order to provide a consistent description of the ionization history and its relations to the \lya emission, which we now describe. 
\subsection{IGM Ionized fraction calculation}
\label{sec:SimFast_xi_Tk}

In the previous version of the SimFast21 code, the IGM ionized fraction was computed assuming that at each redshift the ionization state of a region could be estimated from the collapsed mass in that region assuming a linear relation between collapsed mass and ionizing power. So a given spherical region of radius R is considered ionized if \citep{2006PhR...433..181F}:
\be
\zeta\ M_{\rm coll}(R)  \ge M_{\rm tot}(R),
\ee
where $M_{\rm coll}$ is the collapsed mass which corresponds to the total mass in halos in that region, $M_{\rm tot}$ is the total mass in the region and $\zeta$ is an ionizing efficiency parameter. This efficiency parameter tries to include all the ionizations and recombinations produced by a halo as a function of its mass but has no actual physical meaning although its use is somewhat justified by the large uncertainty in the astrophysical quantities involved in the determination of the relation between halo mass and ionizing efficiency and in the adjustment of this parameter in order to reproduce a reionization history compatible with observations. 

In order to calculate the \lya field however, we need to include the recombinations in the IGM explicitly, as well as directly relate the ionization process to the emitted stellar radiation. We therefore modified the SimFast21 code to include these improvements. 
This new method allows a non linear relation between collapsed mass and ionizing power and all the parameters involved in the calculation have values based in current astrophysical constraints. 
Also, the size of ionized regions is now set by the volume at which the total ionizing emissivity of the sources it contains equals the number of recombinations so that the system is in equilibrium.
For each redshift the implementation of this method was done with the following steps:
\begin{enumerate}

\item A halo catalog with the mass and three dimensional spatial positions was generated using the same method used in the original version of the SimFast21 code.

\item We calculated SFRs from the halo catalogs using the non-linear relations in equations \ref{SFR_param1}, \ref{SFR_param2} and \ref{SFR_param3}.

\item We used equation \ref{eq:Nion} to obtain the halo ionizing rate, $\dot{N}_{\rm ion}$, we corrected for the presence of helium using $A_{\rm He}$ and multiplied it by $f_{\rm esc}$, to account for the photons consumed inside the galaxies.  
\be
\dot{N}_{\rm ion}^{\rm IGM}(z,M)=A_{\rm He}\dot{N}_{\rm ion}(M)f_{\rm esc}(z,M).
\ee
The UV ionizing rates of the halos, $\dot{N}_{\rm ion}^{\rm IGM}$ were then put in three dimensional boxes.

\item Three dimensional boxes with the rate of recombinations in each cell, $\dot{N}_{\rm rec}^{\rm IGM}=V_{cell}\times\dot{n}_{\rm rec}^{\rm IGM}$, were obtained from a dark matter density simulation made with the SimFast21 code using equation \ref{eq:nrec_s} with $x_i$ set to one and $T_K=10^4K$.
\item Following the same procedure as in the original version of the SimFast21 code we applied a series of top-hat filters of decreasing size (this filtering procedure was done in Fourier space) to the ionizing rate and the recombination rate boxes in order to calculate the region ionizing rate and recombination rate.

\item At each filtering step of radius R we found the ionized regions (HII bubbles) by checking if the region ionizing rate was equal or higher than its recombination rate. With this method HII bubbles are always fully ionized:
\be
\dot{N}_{\rm ion}^{\rm IGM}(z,R) \ge \dot{N}_{\rm rec}^{\rm IGM}
\ee

\end{enumerate}

\subsection{Intensity from recombinations and collisions in the IGM}

The SimFast21 code was built to calculate the IGM ionized state assuming two types of regions: one fully ionized (inside the HII bubbles) and other fully neutral. 
The intensity of \lya emission in the IGM due to recombinations is a smooth function of the IGM ionized fraction and is dominated by emission from fully ionized regions (see figure~\ref{I_tk_xi1}) so the output os the SimFast21 code is good enough to estimate this intensity.

Collisions between electrons and neutral hydrogen atoms can also lead to \lya emission, however as was explained in section~\ref{sec:3.2}, collisional \lya emission only occurs in partly ionized regions, mainly in the the edge of HII bubbles, so the estimation of this emission requires a more detailed description of the IGM ionized state than the one given by the limited resolution of semi numerical simulations. 

Collisions are most important in regions where the IGM ionized fraction is locally close to 0.5 and the temperatures are high. Since high temperature regions are likely to be highly ionized we can deduce with the help of figure \ref{I_tk_xi1}, that \lya emission from recombinations is dominant over \lya emission from collisions in the IGM.

\subsection{Intensity from the scattering of Lyman-n photons in the IGM}

The IGM \lya intensity from scattering of \lyn photons emitted from galaxies can also be calculated using data from the code SimFast21. This code uses Equation 10 in \cite{2010MNRAS.406.2421S} to calculate the spherical average of the number of \lya photons, $J_\alpha$, hitting a gas element per unit proper area per unit time per unit frequency per steradian.\\
The Lyman alpha intensity originated from these continuum photons is given by:
\be
I^{\rm IGM}_{\rm cont}=\frac{6 J_{\alpha}E_{{\rm Ly}\alpha}D_{\rm A}^2}{(1+z)^2 D_{\rm L}^2}.\\
\ee

\subsection{Results}
\label{sec:SimFastLya}

Using the prescriptions described in the previous sections we ran simulations \textit{Sim1}, \textit{Sim2} and \textit{Sim3} with a volume of $54^{3}h^{-3}\ \rm Mpc^{3}$ and $1800$ cells from redshift 14 to redshift 6.

The obtained IGM ionization fractions, at redshift seven, where $x_i =$0.86 for simulation \textit{Sim1} and $x_i =$1.0 for simulations \textit{Sim2} and \textit{Sim3}. These values are consistent with the current most likely values for this parameter, $0.8 \le x_i(z=7) \le 1.0$ \citep{2012MNRAS.419.1480M}. 

\begin{figure}[bt!]
\begin{center}
\hspace{-23pt}
\includegraphics[scale = 0.51]{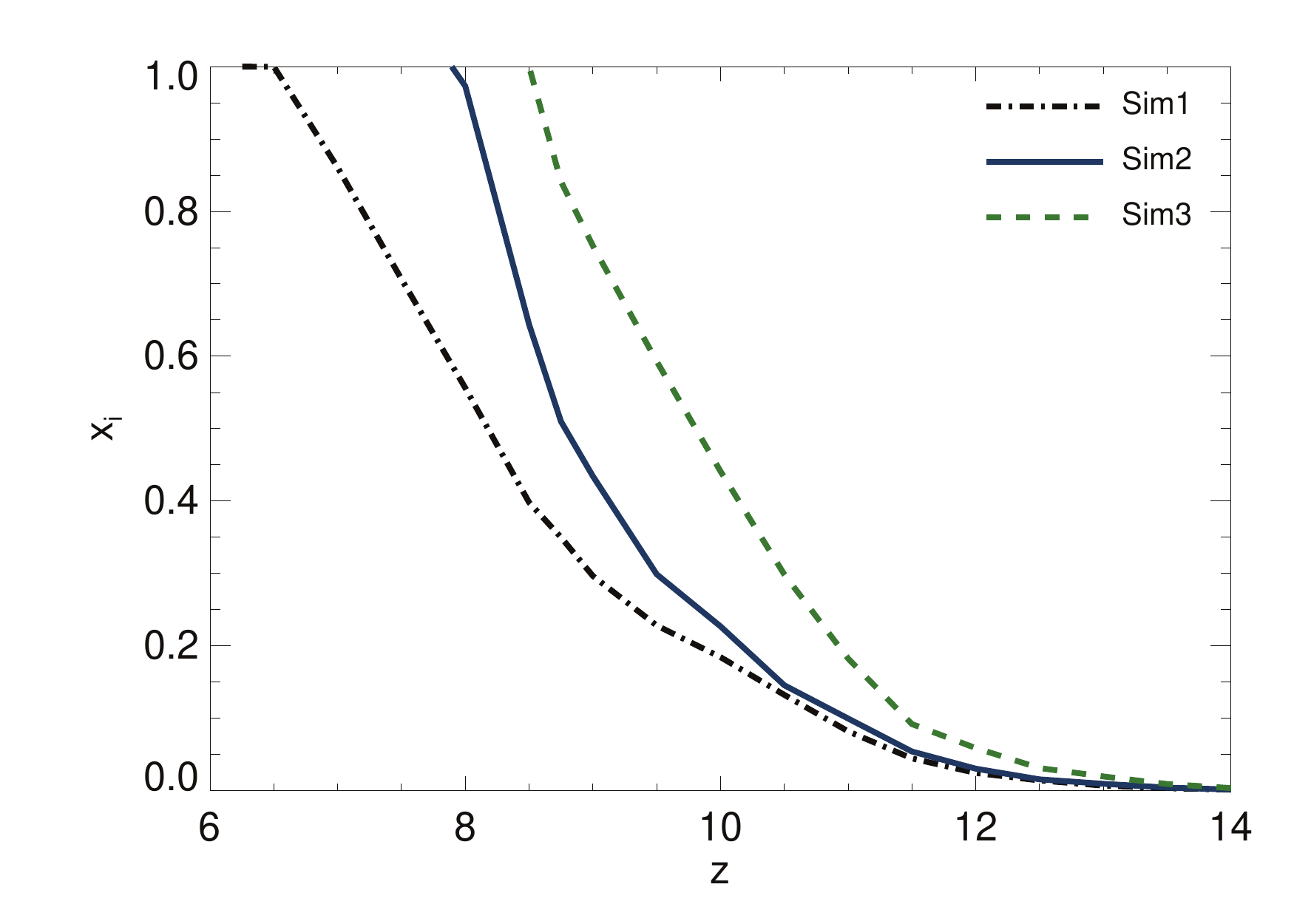} 
\caption{Evolution of the IGM ionized fraction as a function of redshift for the three star formation rate halo mass parameterizations shown in section \ref{sec:SFRvsMhalo}.}
\label{fig:xi_z}
\end{center}
\end{figure}

The IGM ionized fraction evolution for \textit{Sim2} and for \textit{Sim3} (see figure \ref{fig:xi_z}), resulted in optical depths to reionization of $0.073$ and $0.082$. This optical depths are consistent with the value obtained by WMAP ($\tau=0.088\pm0.015$) \citep{2011ApJS..192...18K}. The optical depth correspondent to \textit{Sim1} is 0.66 which is lower than the current observational constraints. Based in the optical depth constraint \textit{Sim2} and \textit{Sim3} have the most likely reionization histories and the IGM ionized fraction evolution obtained with \textit{Sim1} can be seen as a lower bound.
The intensities of Lyman alpha emission from galaxies at redshift seven obtained with the SimFast21 code are similar to the more theoretical estimates summarized in table~\ref{table_GAL}. 

Intensities of Lyman alpha emission in the IGM made with the same code are presented in table~\ref{table_IGM}. 

\begin{table}[h!]
\centering                         
\begin{tabular}{l  c  c  c}        
\hline\hline                 
Source of emission in  & $\nu$I$_\nu$(z=7)  & $\nu$I$_\nu$(z=8) & $\nu$I$_\nu$(z=10) \\
$[{\rm erg s^{-1} cm}^{\rm -2}{\rm sr}^{\rm -1}]$ \\   
\hline                      
   Recombinations & $9.3\times10^{-10}$ & $4.8\times10^{-10}$ & $9.6\times10^{-11}$\\ 
   Continuum      & $3.5\times10^{-13}$ & $1.2\times10^{-13}$ & $1.5\times10^{-14}$\\   
   Total          & $1.6\times10^{-9}$ & $6.7\times10^{-10}$ & $1.1\times10^{-10}$\\

\hline                                  
\end{tabular}
\caption{Surface brightness (in observed frequency times intensity) of \lya emission from the different sources in the IGM at $z\approx 7$, $z\approx 8$  and $z\approx 10$.}
\label{table_IGM}     
\end{table}

The intensity values found in tables \ref{table_GAL} and \ref{table_IGM} and the theoretical estimations plotted in figure \ref{I_tk_xi2} indicate that for the \lya intensity from the IGM to reach a value close to the emission from galaxies at $z=7$ would require a very large absorption of \lya photons by dust in galaxies. 

\begin{figure}[htbp]
\begin{center}

\hspace{-16pt}
\includegraphics[scale = 0.51]{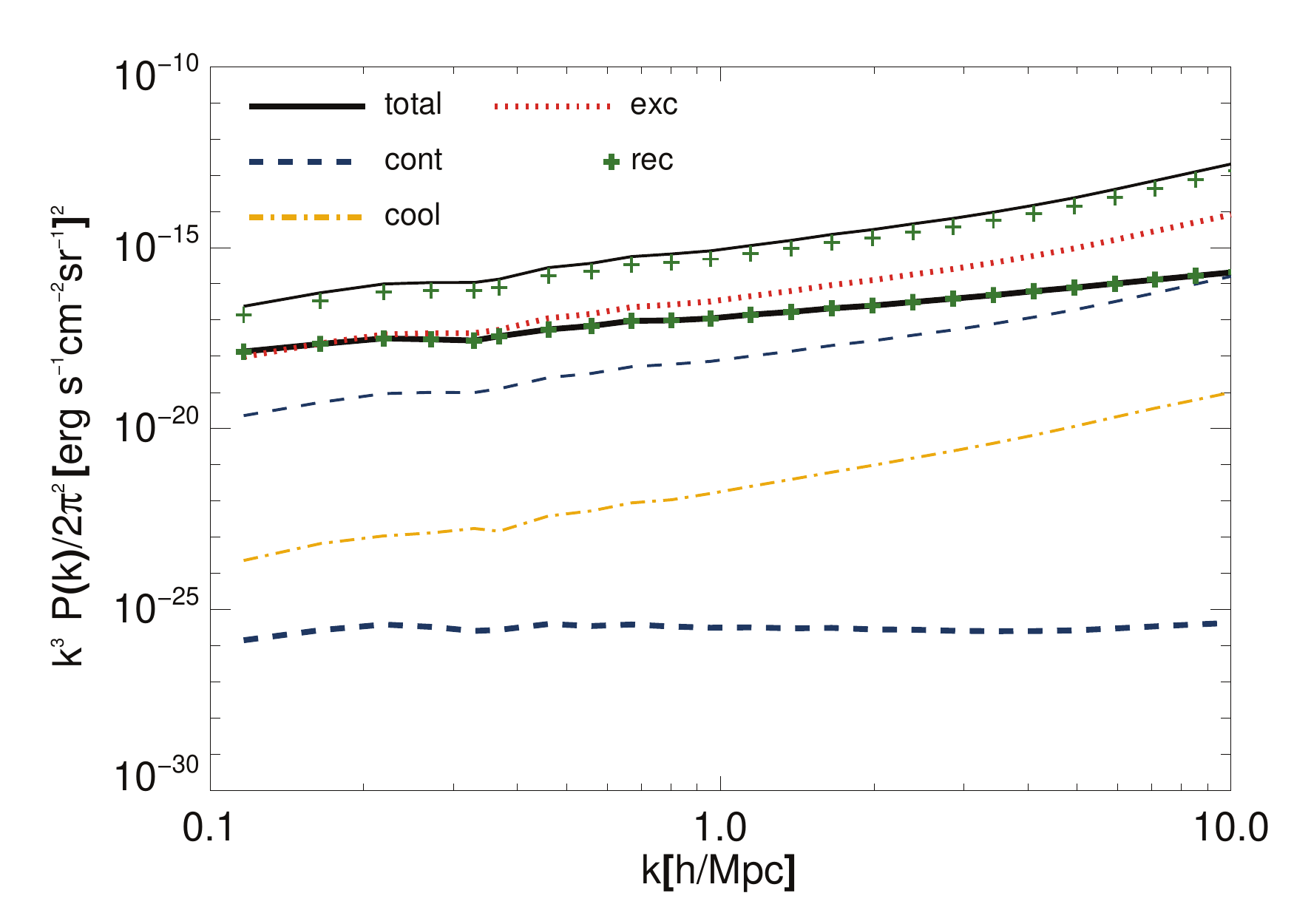}

\hspace{-16pt}
\includegraphics[scale = 0.51]{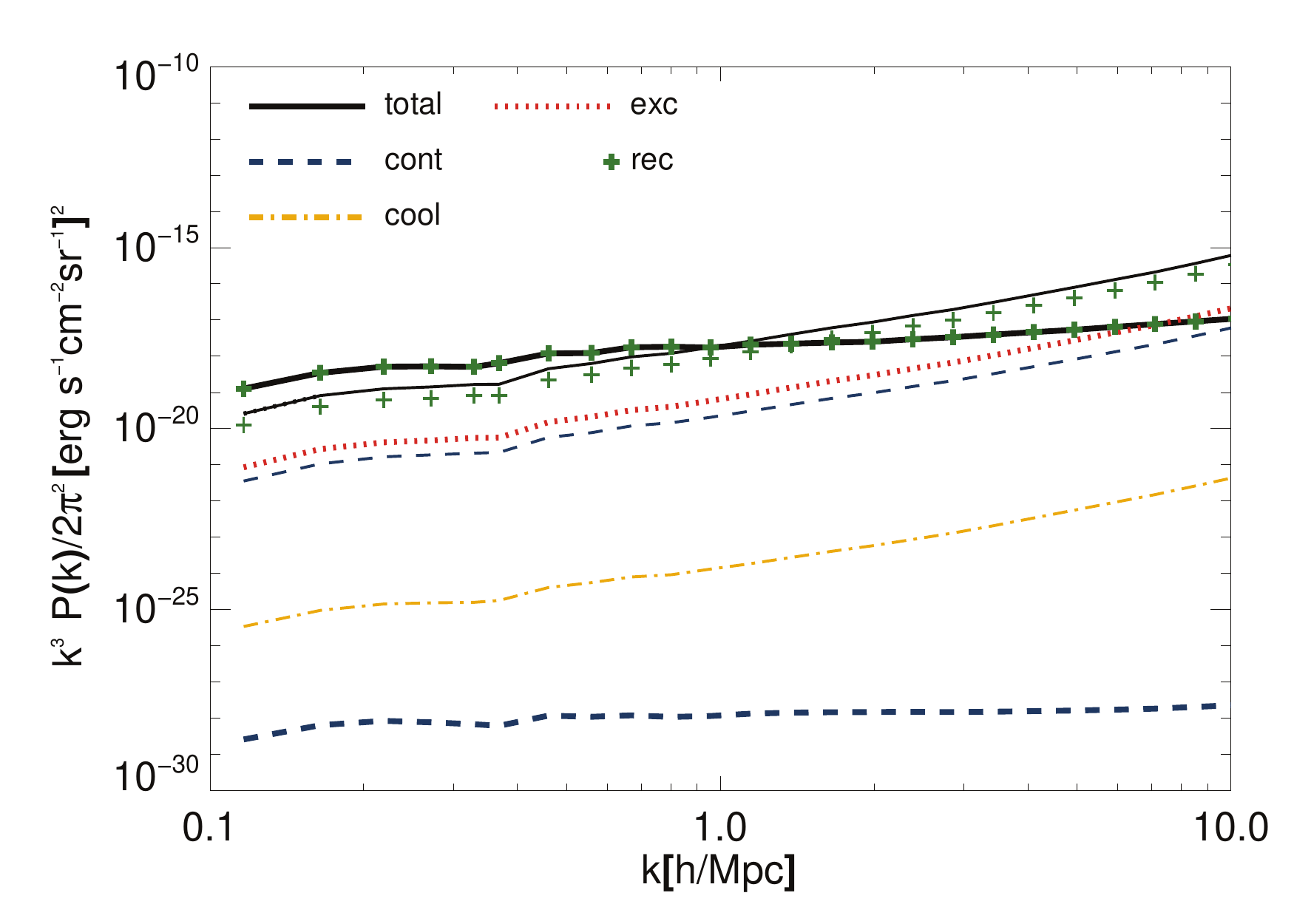}
\end{center}
\caption{Power spectrum of \lya surface brightness ($\nu I$) from galaxies (thin lines) and from the IGM (thick lines) at redshifts 7 (top) and 10 (bottom). The shown contributions to the \lya flux are from: excitations and collisions, recombinations, continuum emission inside the \lya width (from galaxies), scattering of \lyn photons (from the IGM), cooling emission in galaxies and total emission.}
\label{fig:ps_Lya_z7_z10}
\end{figure}

The resulting power spectra of \lya emission in galaxies and in the IGM obtained with the SimFast21 code are presented in figure~\ref{fig:ps_Lya_z7_z10} for $z=7$ and for $z=10$.

\begin{figure}[htbp]
\begin{center}
\hspace{-16pt}
\includegraphics[scale = 0.51]{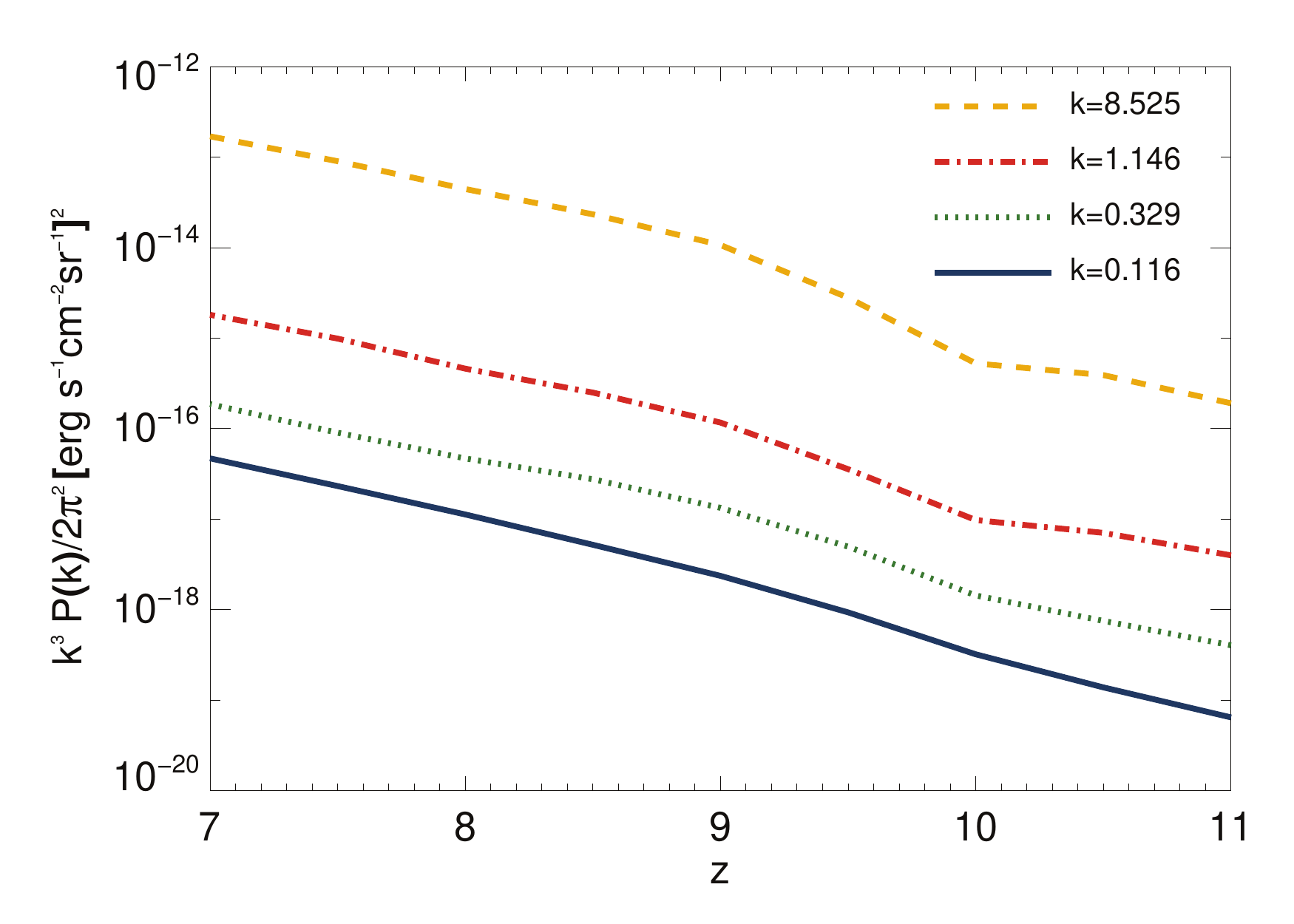} 
\caption{Total power spectrum of \lya emission during the EOR as a function of redshift.}
\label{fig:Lya_tot_k}
\end{center}
\end{figure}

We repeated the \lya power spectra calculation for several redshifts during the EOR and plotted the \lya power spectra as a function of redshift for several $k$ in figure~\ref{fig:Lya_tot_k}.

\begin{figure}[htbp]
\begin{center}
\hspace{-16pt}
\includegraphics[scale = 0.51]{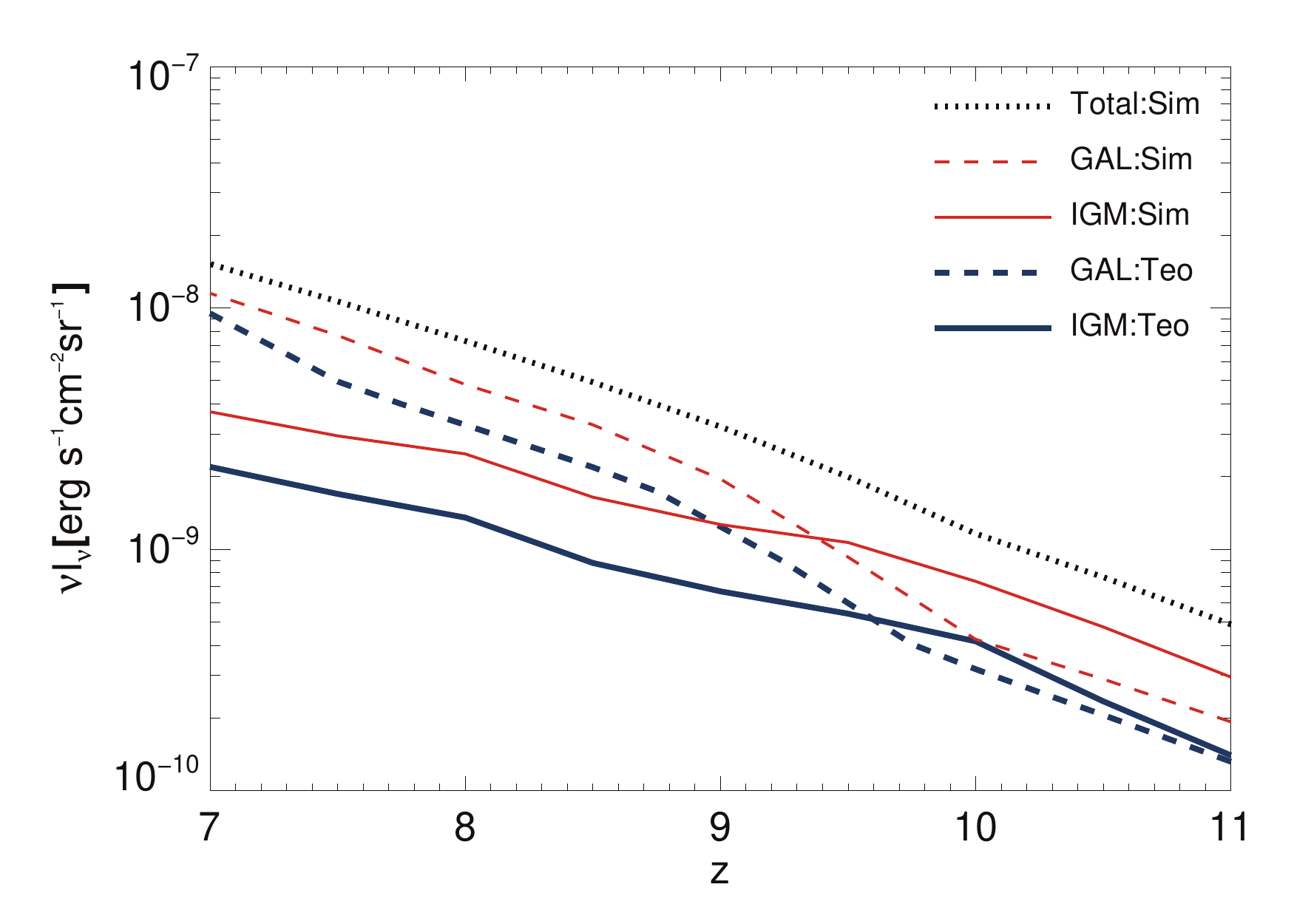} 
\caption{\lya Intensity from galaxies (dashed lines), from the IGM (solid lines) as a function of redshift from our simulation (red thin lines) and from the theoretical calculations (blue thick lines). Also shown is the total \lya emission from the simulation (dotted line). All ‎the intensities where calculated using the star formation halo mass relation from \textit{Sim1}. The theoretical intensity of \lya emission from the IGM was calculated using the average IGM ionization values obtained from \textit{Sim1}.}
\label{fig:Int_IGM_GAL}
\end{center}
\end{figure}

We calculated the intensity of \lya emission from galaxies and from the IGM (intensities are shown in figure~\ref{fig:Int_IGM_GAL}), and found that according with our assumptions and as already previously seen, the \lya emission from galaxies is dominant over the \lya emission from the IGM at least during the redshift interval from $z=6$ to $z=9$.

Since the star formation halo mass relation is not very constrained, we can use the results obtained with \textit{Sim1} and \textit{Sim3} as the lower and upper bounds to the expected \lya intensity.

The evolution of the \lya intensity from galaxies, from the IGM and total intensity can be seen respectively in tables \ref{table_GAL_sim} \ref{table_IGM_sim} and \ref{table_total_sim} for simulations \textit{Sim1}, \textit{Sim2} and \textit{Sim3}.

\begin{table}
\centering                         
\begin{tabular}{l  c  c  c}        
\hline\hline                 
Simulation   & $\nu$I$_\nu$(z=7)  & $\nu$I$_\nu$(z=8) & $\nu$I$_\nu$(z=10) \\   
\hline                      
   \textit{Sim1} & $1.43\times10^{-8}$ & $5.13\times10^{-9}$ & $4.55\times10^{-11}$\\ 
   \textit{Sim2} & $2.54\times10^{-8}$ & $8.34\times10^{-9}$ & $5.68\times10^{-11}$\\   
   \textit{Sim3} & $3.57\times10^{-8}$ & $1.26\times10^{-8}$ & $9.73\times10^{-11}$\\

\hline                                  
\end{tabular}
\caption{Surface brightness (in observed frequency times intensity) in units of $[{\rm erg s^{-1} cm}^{\rm -2}{\rm sr}^{\rm -1}]$ of \lya emission from Galaxies at $z\approx 7$, $z\approx 8$ and $z\approx 10$ for \textit{Sim1}, \textit{Sim2} and \textit{Sim3}.}
\label{table_GAL_sim}     
\end{table}


\begin{table}
\centering                         
\begin{tabular}{l  c  c  c}        
\hline\hline                 
Simulation   & $\nu$I$_\nu$(z=7)  & $\nu$I$_\nu$(z=8) & $\nu$I$_\nu$(z=10) \\
\hline                      
   \textit{Sim1} & $4.33\times10^{-9}$ & $3.76\times10^{-9}$ & $1.74\times10^{-9}$\\ 
   \textit{Sim2} & $6.07\times10^{-9}$ & $5.17\times10^{-9}$ & $2.18\times10^{-9}$\\   
   \textit{Sim3} & $8.53\times10^{-9}$ & $7.81\times10^{-9}$ & $3.76\times10^{-9}$\\

\hline                                  
\end{tabular}
\caption{Surface brightness (in observed frequency times intensity) in units of $[{\rm erg s^{-1} cm}^{\rm -2}{\rm sr}^{\rm -1}]$ of \lya emission from the IGM at $z\approx 7$, $z\approx 8$ and $z\approx 10$ for \textit{Sim1}, \textit{Sim2} and \textit{Sim3}.}
\label{table_IGM_sim}     
\end{table}

\begin{table}
\centering                         
\begin{tabular}{l  c  c  c}        
\hline\hline                 
Simulation   & $\nu$I$_\nu$(z=7)  & $\nu$I$_\nu$(z=8) & $\nu$I$_\nu$(z=10) \\   
\hline                      
   \textit{Sim1} & $1.86\times10^{-8}$ & $8.89\times10^{-9}$ & $1.79\times10^{-9}$\\ 
   \textit{Sim2} & $3.15\times10^{-8}$ & $1.35\times10^{-8}$ & $2.24\times10^{-9}$\\   
   \textit{Sim3} & $4.42\times10^{-8}$ & $2.04\times10^{-8}$ & $3.86\times10^{-9}$\\

\hline                                  
\end{tabular}
\caption{Surface brightness (in observed frequency times intensity) in units of $[{\rm erg s^{-1} cm}^{\rm -2}{\rm sr}^{\rm -1}]$ of total \lya emission at $z\approx 7$, $z\approx 8$ and $z\approx 10$ for \textit{Sim1}, \textit{Sim2} and \textit{Sim3}.}
\label{table_total_sim}     
\end{table}

\begin{figure}[htbp]
\hspace{-15pt}
\includegraphics[scale = 0.52]{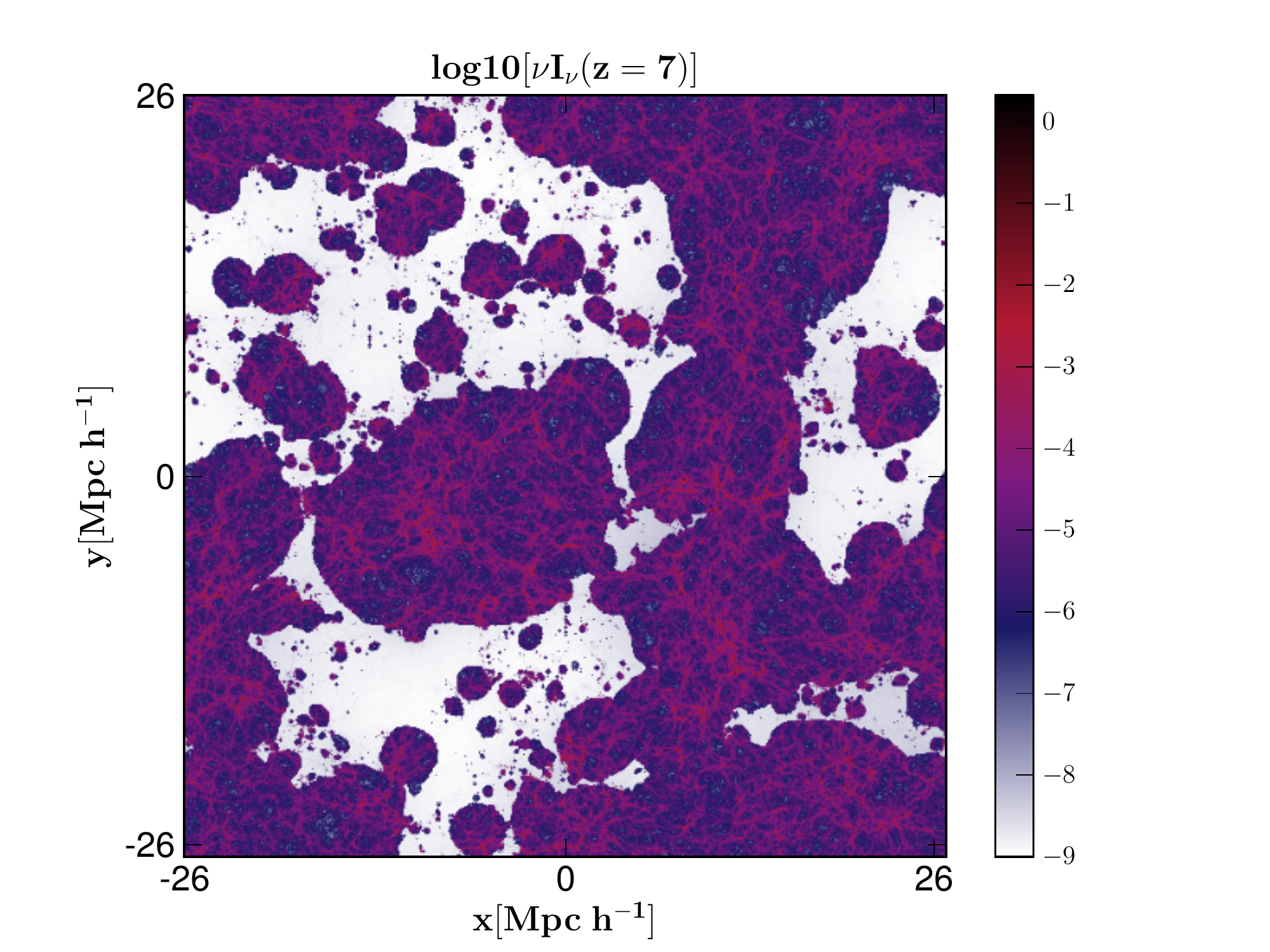}

\caption{Total \lya Intensity from galaxies and the IGM in erg s$^{-1}$ cm$^{-2}$ sr$^{-1}$ at redshift 7.}
\label{fig:map_Lya_Total}
\end{figure}

A map of the total \lya intensity in galaxies and in the IGM is presented in figure~\ref{fig:map_Lya_Total} for $z=7$. 

\section{Cross-correlation between \lya and 21-cm observations}
\label{sec:cross_correlation}

Observations of the 21 cm signal from the EOR will suffer from contamination by foregrounds and systematic effects. Since both 21 cm line emission and \lya line emission trace neutral hydrogen, these two lines are expected to be strongly correlated. The cross correlation of these two lines can be used as an extra method to probe the evolution of the IGM ionized hydrogen fraction. In particular the power spectra of this cross correlation will have a discontinuity at a scale that is related to the average bubble size and hence the average ionization fraction in the Universe.

During the EoR, the 21 cm signal from galaxies is much smaller than the emission from the IGM so it is safe to neglect both this galaxy emission and the shot noise emission in the cross-correlation. Since the \lya emission from galaxies is dominating over the IGM for most redshifts, we can just concentrate on the \lya-galaxy/21cm-IGM cross-correlation when analyzing the cross-power spectrum.
The cross correlation between the 21 cm signal and the \lya line in galaxies is therefore given by:
\be
P_{\rm Ly \alpha, 21}(z,k)=I_{\rm GAL} I_{\rm 21} \left[ P_{\delta \delta}-\frac{1}{1-\bar{x}_i}P_{x_i \delta} \right]\,
\label{eq:ps_cross_gal}
\ee
where $I_{\rm 21}$ is the average intensity of 21 cm emission, $P_{x_i \delta}(z,k)$ is the cross correlation power spectra between the ionized field and the matter density fluctuations, $P_{\delta\delta}(z,k)$ is the power spectra of matter density fluctuations and we are assuming that the \lya emission is a biased tracer of the underlying dark matter field.

\begin{figure}[t!]
\begin{center}
\hspace{-16pt}
\includegraphics[scale = 0.5]{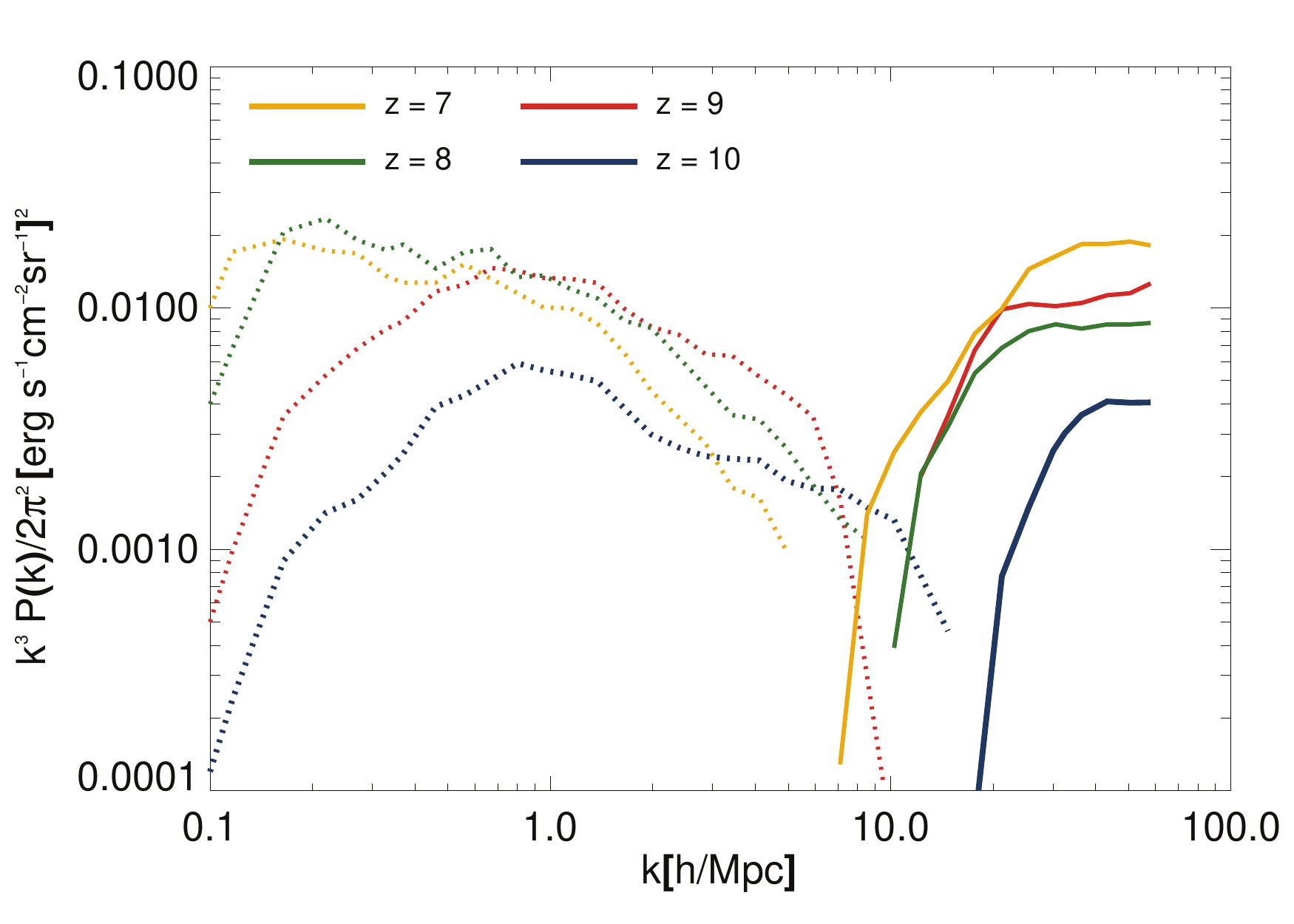} 
\caption{Cross correlation power spectrum between \lya emission and the 21 cm signal for redshifts $7$, $8$, $9$ and $10$. Dotted lines indicate a negative correlation and solid lines indicate a positive correlation.}
\label{fig:ps_cross}
\end{center}
\end{figure}

In figure~\ref{fig:ps_cross} we show the cross-correlation power spectrum between the total \lya emission and the 21 cm signal at redshifts $7$, $8$, $9$ and $10$. For simulation \textit{Sim1} this redshifts correspond to ionizing fractions of $x_i= 0.86$, $x_i= 0.56$, $x_i= 0.35$ and $x_i= 0.23$ for redshifts $7$, $8$, $9$ and $10$ respectively.
In figure~\ref{fig:ps_cross} the scale at which $P_{\rm Ly \alpha, 21}(k)$ goes from negative to positive is determined by the average size of the ionized regions. For small scales the correlation is positive since fluctuations from both lines should be proportional to the underlying density fluctuations but for large scales (small $k$) the correlation is negative since the 21 cm line and the lyman alpha line are characteristic of neutral gas and ionized gas respectively (and there will be an extra negative contribution from the ionised bubbles).

\section{Observations}
\label{sec:observation}

Current observations related to \lya emission are restricted to narrow-band imaging of \lya emitters during reionization and the direct detection of individual emitters. This has led to order $\sim 200$ secure detections at $z > 5$, but systematic uncertainties remain on the fraction that are arising at low redshifts and associated with [OIII]/[OII] lines, among others. Due to narrow atmospheric windows, observations in near-IR bands necessary to cover the reionization epoch are also limited to multiple discrete bands. In any case, existing data could be used for a statistical study such as the power spectrum to extract properties of \lya emitters that remain below the 5$\sigma$ level of individual source/line identifications. Given that detections do exist at the bright-end and our predictions are consistent with the \lya LFs derived from observational measurements, it is likely that a modest improvement in existing technology and programs will lead to an experiment with sufficient sensitivity to measure 
the 
\lya anisotropy power spectrum during reionization over a broad range of redshifts. The main limitation, unfortunately, is that existing ground-based observations are very limited to small fields of view with narrow-bands in the redshift.

Note that from the ground we expect a noise ($\nu I$) of $\sim 2.5\times10^{-3}$ erg cm$^{-2}$ sr$^{-1}$ (assuming we can avoid the OH lines, otherwise, the intensity will be $\sim 1.0\times10^{-1}$ erg cm$^{-2}$ sr$^{-1}$). From space, the main contamination will be the zodiacal light, which will have a value $\sim 5\times10^{-4}$ erg cm$^{-2}$ sr$^{-1}$. It is possible that a dedicated experiment from the ground can be conceived to improve our understanding of reionization through detailed \lya mapping over a broad range of redshifts using specific instruments and filters that suppress the atmospheric contamination. Because of this strong atmospheric contamination, sub-orbital and/or orbital experiments may however offer a better option. The predictions we have made here can be used as a guide in designing such instruments and experiments.  

\begin{figure}[t!]
\begin{center}
\hspace{-16pt}
\includegraphics[scale = 0.51]{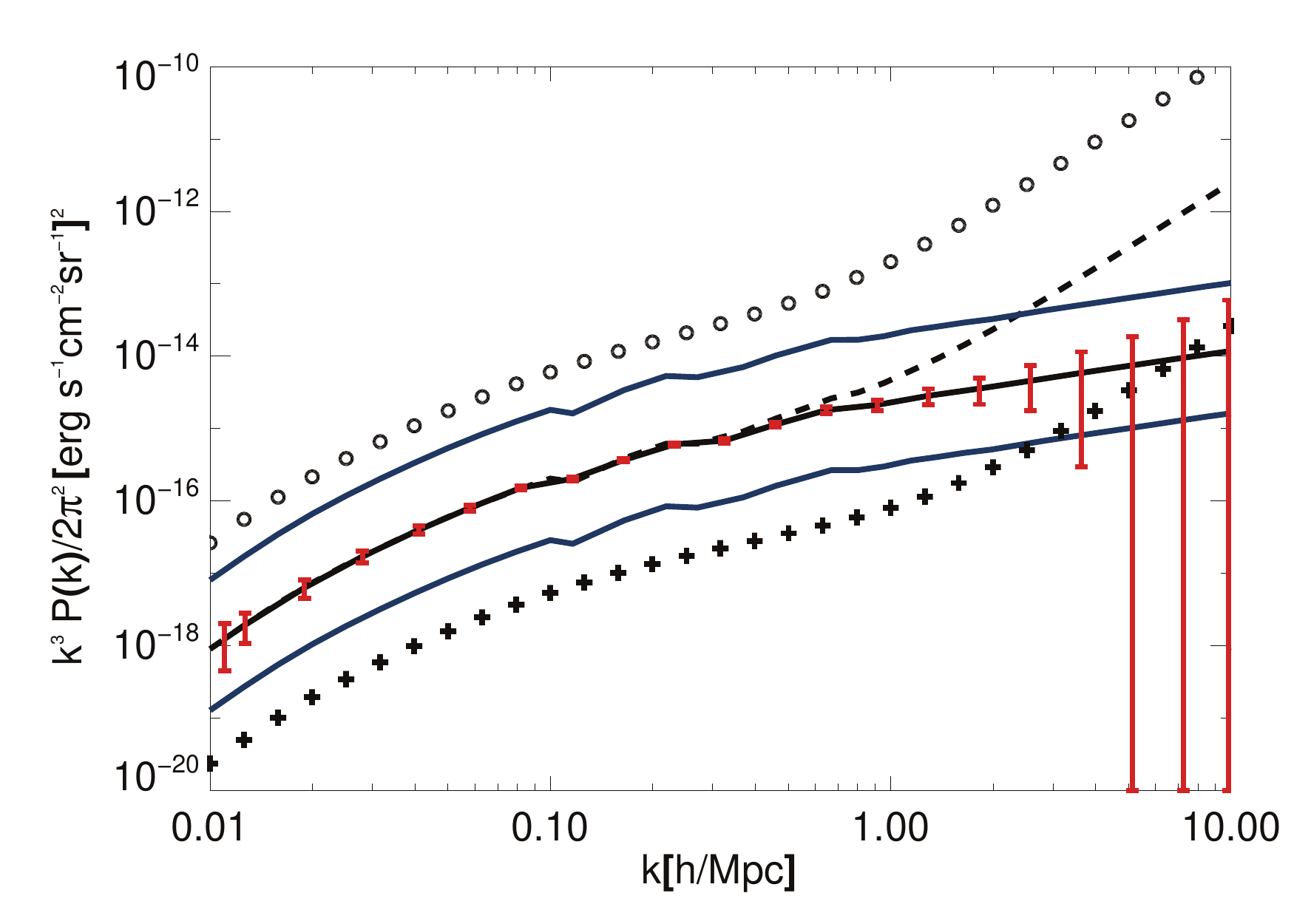} 
\caption{Expected error on the \lya clustering power spectrum at $z=7$ using a space based experiment. Black solid line shows the clustering power spectrum for {\it Sim3} while the dashed line includes the shot noise. Red vertical bars shows the error. The lower blue solid line shows the clustering power spectrum for {\it Sim1} while the top blue solid line shows the same for a model similar to {\it Sim3} with the same reionization history and optical depth (from WMAP) but with a SFR 3 times larger and a UV escape fraction 3 times lower, which will generate a \lya luminosity function larger than what is usually expected at $z=7$. Circles show the expected H$\alpha$ power spectrum from $z=0.5$ that will contaminate the observation and the crosses gives the expected H$\alpha$ signal after galaxies with a H$\alpha$ luminosity $> 1.0\times 10^{40}$ erg s$^{-1}$ are removed.}
\label{fig:ps_err}
\end{center}
\end{figure}

In figure \ref{fig:ps_err} we show the expected errors at $z=7$ (central wavelength of 0.975 $\mu$m) for a dedicated compact space-borne template instrument to study \lya EOR fluctuations.  We consider a 20 cm aperture and a spectrometer with resolution $R = \lambda/\Delta \lambda = 200$. The imaging will be done using a 2048x2048 HgCdTe detector array in order to cover in one pointing a field of view of 45x45 arcmin with a resolution of 10 arc-second pixels on the sky and a spectral range from 0.85 to 1.10 $\mu$m.
We took a survey area of 20 deg$^2$ and a total observation time of 2900 hours. This example shows that \lya EOR science is well within the reach of our modest template instrument.  The calculated sensitivities achieved on the deep fields are sufficient to detect \lya in broad $\Delta k/k$ bins ranging from $k = 0.01$ to $10$ h/Mpc in both clustering and Poisson fluctuations.

Ideally the spectral resolution would match the maximum k available in the angular direction; however higher spectral resolution requires longer integration times needed to realize photon noise limited sensitivity, which tends to degrade the instrument sensitivity. The angular resolution does not affect surface brightness sensitivity directly, but does determine the depth to which lower-redshift galaxies may be masked using a deep ancillary continuum galaxy survey. Although the continuum emission from galaxies can in principle be removed by looking at the signal across the frequency direction, as explained before, contamination from other lines at lower redshifts does poses a problem to the detection of the \lya signal, in particular from the H$\alpha$ line. The most straightforward way to remove this contamination would be by masking the pixels where these low-z galaxies are found, either from the observation itself or using another, high sensitivity, continuum observation. For this approach, the angular 
resolution of the \lya experiment has to be good enough in order to have enough pixels left after the masking.
Therefore, this instrument is required to have higher angular resolution than spectral resolution. 
Figure \ref{fig:ps_err} also shows the expected contamination from the H$\alpha$ line from galaxies at $z=0.5$ (black dots). This was calculated following the same approach as for the \lya line and using the H$\alpha$ to SFR relation taken from \citet{1994ApJ...435...22K} and \citet{1998ApJ...498..541K}.
Removing low-z galaxies down to a mass of $\sim 6.6\ 10^{10}$ M$_\sun$, corresponding to a cut in Luminosity $L>1.0\times 10^{40}$ erg/s, would bring this contamination  below the \lya signal (black crosses). Using the H$\alpha$ Luminosity function from \citet{2010MNRAS.402.1330G} normalized to the SFR density at $z=0.5$ we get an expected angular density of about 25 H$\alpha$ emitters per square degree per band, which would mean that only $\sim$ 0.98\% of the pixels would be masked.

Note that the rejection of interloping low-redshift galaxies requires a full treatment that is beyond the scope of this paper.  Foreground rejection may also be significantly enhanced by simultaneously detecting additional EOR spectral features beyond \lya, which are produced by interlopers with very low probability. Combining these \lya measurements with other EOR observations (CO, C+ and particularly HI 21 cm) offers additional information on EOR star-formation, metallicity, and ionization history. The possibility of constructing an experiment in a near-IR band to measure the \lya flux in order to correlate it with the 21 cm signal was also explored by \cite{2007arXiv0705.1825W}. Although, they used simple models to estimate the fluctuations in each of these two lines, they also considered several foregrounds that will contaminate the observations and concluded that it is possible to remove enough foregrounds that the intensity of radiation emitted from galaxies can be constrained from the cross 
correlation.\\

\section{Summary}
\label{sec:discussion}

In this paper we took into account the main contributions to \lya emission from recombinations, collisions, continuum emission in galaxies and scattering of Lyman-n photons to calculate the intensity of 
\lya emission from galaxies and from the IGM during the EOR.

We started by theoretically calculating the intensities using astrophysical data from several observational results and then implemented the calculation in a simulation using a modified version of the code SimFast21 to obtain the spatial fluctuations of \lya emission. The simulation allowed to calculate the \lya emission taking into account the spatial fluctuations of the different astrophysical parameters, which represents an improvement over theoretical calculations that only use the average values. 

Our simulations showed that to achieve optical depths compatible with the WMAP constraints the high SFRD required imply that for reasonable values of UV and \lya escape fraction the intensity of \lya emission from galaxies is dominant over the emission from the IGM. 

By testing different SFR halo mass parameterizations we constrained the intensity of \lya emission from galaxies to be about $(1.43-3.57)\times 10^{-8}$ and $(4.55-9.73)\times10^{-11}$ erg $^{-1}$ cm$^{-2}$ sr$^{-1}$ at redshift 7 and 10, respectively which is dominant over the intensity of \lya emission from the IGM at $z=7$ (about $1.6\times 10^{-5}$) but less at $z=10$ ($1.1\times10^{-10}$ erg s$^{-1}$ cm$^{-2}$ sr$^{-1}$).  Since the intensity levels we found are lower than the extragalactic background intensity from galaxies and so are too low to be detected with an experiment aiming the absolute background intensity, we propose an intensity mapping experiment which will allow to measure the \lya power spectrum.

For reasonable astrophysical conditions the process of hydrogen reionization was done by UV radiation originated in galaxies with luminosities below the high redshift observational threshold. In this work we showed the different ways by which UV emission is connected to \lya emission and so we stress how it would be useful to use intensity mapping of \lya emission to probe the overall intensity of UV radiation.  

\lya emission can also be connected to the 21 cm signal from the Epoch of Reionization, since the continuum photons above the \lya line that redshift to this line in the IGM contribute to the radiative coupling of the 21 cm signal to the gas temperature. 
The cross correlation of the \lya and the 21-cm lines can be used to reduce systematics and foregrounds encountered with 21-cm observations. In particular the discontinuity of the cross correlation power spectra will provide constrains in the evolution of the IGM ionized fraction.

In previous studies we have discussed the use of CO molecular and CII fine-structure atomic lines to complement 21-cm data in the attempt to probe the IGM during reionization. Our study shows that \lya intensity mapping is also a viable approach to probe reionization and is within the experimental reach over the coming decade.\\

\begin{acknowledgments}

This work was supported by FCT-Portugal with the grant (SFRH/BD/51373/2011) for MBS and  
under grant PTDC/FIS/100170/2008 for MBS and MGS.\\ 
AC and YG acknowledge support from NSF CAREER AST-0645427 and NASA NNX10AD42G at UCI.\\
MBS was a long-term Visiting Student at UCI, supported by NSF CAREER AST-0645427, 
when this work was initiated and she thanks the Department of Physics and Astronomy at UCI for hospitality during her stay.\\

\end{acknowledgments}

\bibliographystyle{apj}
\bibliography{apj-jour,Lyalpha}

\begin{thebibliography}{84}
\expandafter\ifx\csname natexlab\endcsname\relax\def\natexlab#1{#1}\fi

\bibitem[{{Barkana} \& {Loeb}(2001)}]{2001PhR...349..125B}
{Barkana}, R., \& {Loeb}, A. 2001, \physrep, 349, 125

\bibitem[{{Barkana} \& {Loeb}(2005)}]{2005ApJ...626....1B}
---. 2005, \apj, 626, 1

\bibitem[{{Bouwens} {et~al.}(2011{\natexlab{a}}){Bouwens}, {Illingworth},
  {Labbe}, {Oesch}, {Trenti}, {Carollo}, {van Dokkum}, {Franx}, {Stiavelli},
  {Gonz{\'a}lez}, {Magee}, \& {Bradley}}]{2011Natur.469..504B}
{Bouwens}, R.~J., {Illingworth}, G.~D., {Labbe}, I., {et~al.}
  2011{\natexlab{a}}, \nat, 469, 504

\bibitem[{{Bouwens} {et~al.}(2011{\natexlab{b}}){Bouwens}, {Illingworth},
  {Oesch}, {Trenti}, {Labbe}, {Franx}, {Stiavelli}, {Carollo}, {van Dokkum}, \&
  {Magee}}]{2011arXiv1105.2038B}
{Bouwens}, R.~J., {Illingworth}, G.~D., {Oesch}, P.~A., {et~al.}
  2011{\natexlab{b}}, ArXiv e-prints

\bibitem[{{Bouwens} {et~al.}(2011{\natexlab{c}}){Bouwens}, {Illingworth},
  {Oesch}, {Franx}, {Labbe}, {Trenti}, {van Dokkum}, {Carollo}, {Gonzalez}, \&
  {Magee}}]{2011arXiv1109.0994B}
---. 2011{\natexlab{c}}, ArXiv e-prints

\bibitem[{{Boylan-Kolchin} {et~al.}(2009){Boylan-Kolchin}, {Springel}, {White},
  {Jenkins}, \& {Lemson}}]{2009MNRAS.398.1150B}
{Boylan-Kolchin}, M., {Springel}, V., {White}, S.~D.~M., {Jenkins}, A., \&
  {Lemson}, G. 2009, \mnras, 398, 1150

\bibitem[{{Cantalupo} {et~al.}(2008){Cantalupo}, {Porciani}, \&
  {Lilly}}]{2008ApJ...672...48C}
{Cantalupo}, S., {Porciani}, C., \& {Lilly}, S.~J. 2008, \apj, 672, 48

\bibitem[{{Chen} \& {Miralda-Escud{\'e}}(2008)}]{2008ApJ...684...18C}
{Chen}, X., \& {Miralda-Escud{\'e}}, J. 2008, \apj, 684, 18

\bibitem[{{Chuzhoy} \& {Zheng}(2007)}]{2007ApJ...670..912C}
{Chuzhoy}, L., \& {Zheng}, Z. 2007, \apj, 670, 912

\bibitem[{{Conroy} \& {Wechsler}(2009)}]{2009ApJ...696..620C}
{Conroy}, C., \& {Wechsler}, R.~H. 2009, \apj, 696, 620

\bibitem[{{Cooray} {et~al.}(2012){Cooray}, {Gong}, {Smidt}, \&
  {Santos}}]{2012ApJ...756...92C}
{Cooray}, A., {Gong}, Y., {Smidt}, J., \& {Santos}, M.~G. 2012, \apj, 756, 92

\bibitem[{{Dav{\'e}} {et~al.}(2001){Dav{\'e}}, {Cen}, {Ostriker}, {Bryan},
  {Hernquist}, {Katz}, {Weinberg}, {Norman}, \& {O'Shea}}]{2001ApJ...552..473D}
{Dav{\'e}}, R., {Cen}, R., {Ostriker}, J.~P., {et~al.} 2001, \apj, 552, 473

\bibitem[{{Dayal} {et~al.}(2010){Dayal}, {Ferrara}, \&
  {Saro}}]{2010MNRAS.402.1449D}
{Dayal}, P., {Ferrara}, A., \& {Saro}, A. 2010, \mnras, 402, 1449

\bibitem[{{De Lucia} \& {Blaizot}(2007)}]{2007MNRAS.375....2D}
{De Lucia}, G., \& {Blaizot}, J. 2007, \mnras, 375, 2

\bibitem[{{Dijkstra} {et~al.}(2006{\natexlab{a}}){Dijkstra}, {Haiman}, \&
  {Spaans}}]{2006ApJ...649...14D}
{Dijkstra}, M., {Haiman}, Z., \& {Spaans}, M. 2006{\natexlab{a}}, \apj, 649, 14

\bibitem[{{Dijkstra} {et~al.}(2006{\natexlab{b}}){Dijkstra}, {Haiman}, \&
  {Spaans}}]{2006ApJ...649...37D}
---. 2006{\natexlab{b}}, \apj, 649, 37

\bibitem[{{Dopita} \& {Sutherland}(2003)}]{2003adu..book.....D}
{Dopita}, M.~A., \& {Sutherland}, R.~S. 2003, {Astrophysics of the diffuse
  universe}

\bibitem[{{Fan} {et~al.}(2006){Fan}, {Strauss}, {Becker}, {White}, {Gunn},
  {Knapp}, {Richards}, {Schneider}, {Brinkmann}, \&
  {Fukugita}}]{2006AJ....132..117F}
{Fan}, X., {Strauss}, M.~A., {Becker}, R.~H., {et~al.} 2006, \aj, 132, 117

\bibitem[{{Fardal} {et~al.}(2001){Fardal}, {Katz}, {Gardner}, {Hernquist},
  {Weinberg}, \& {Dav{\'e}}}]{2001ApJ...562..605F}
{Fardal}, M.~A., {Katz}, N., {Gardner}, J.~P., {et~al.} 2001, \apj, 562, 605

\bibitem[{{Fernandez} \& {Komatsu}(2006)}]{2006ApJ...646..703F}
{Fernandez}, E.~R., \& {Komatsu}, E. 2006, \apj, 646, 703

\bibitem[{{Fern{\'a}ndez-Soto} {et~al.}(2003){Fern{\'a}ndez-Soto}, {Lanzetta},
  \& {Chen}}]{2003MNRAS.342.1215F}
{Fern{\'a}ndez-Soto}, A., {Lanzetta}, K.~M., \& {Chen}, H.-W. 2003, \mnras,
  342, 1215

\bibitem[{{Forero-Romero} {et~al.}(2011){Forero-Romero}, {Yepes},
  {Gottl{\"o}ber}, {Knollmann}, {Cuesta}, \& {Prada}}]{2011MNRAS.415.3666F}
{Forero-Romero}, J.~E., {Yepes}, G., {Gottl{\"o}ber}, S., {et~al.} 2011,
  \mnras, 415, 3666

\bibitem[{{Furlanetto} {et~al.}(2006){Furlanetto}, {Oh}, \&
  {Briggs}}]{2006PhR...433..181F}
{Furlanetto}, S.~R., {Oh}, S.~P., \& {Briggs}, F.~H. 2006, \physrep, 433, 181

\bibitem[{{Geach} {et~al.}(2010){Geach}, {Cimatti}, {Percival}, {Wang},
  {Guzzo}, {Zamorani}, {Rosati}, {Pozzetti}, {Orsi}, {Baugh}, {Lacey},
  {Garilli}, {Franzetti}, {Walsh}, \& {K{\"u}mmel}}]{2010MNRAS.402.1330G}
{Geach}, J.~E., {Cimatti}, A., {Percival}, W., {et~al.} 2010, \mnras, 402, 1330

\bibitem[{{Gnedin} {et~al.}(2008){Gnedin}, {Kravtsov}, \&
  {Chen}}]{2008ApJ...672..765G}
{Gnedin}, N.~Y., {Kravtsov}, A.~V., \& {Chen}, H.-W. 2008, \apj, 672, 765

\bibitem[{{Gong} {et~al.}(2012){Gong}, {Cooray}, {Silva}, {Santos}, {Bock},
  {Bradford}, \& {Zemcov}}]{2012ApJ...745...49G}
{Gong}, Y., {Cooray}, A., {Silva}, M., {et~al.} 2012, \apj, 745, 49

\bibitem[{{Gong} {et~al.}(2011){Gong}, {Cooray}, {Silva}, {Santos}, \&
  {Lubin}}]{2011ApJ...728L..46G}
{Gong}, Y., {Cooray}, A., {Silva}, M.~B., {Santos}, M.~G., \& {Lubin}, P. 2011,
  \apjl, 728, L46

\bibitem[{{Gould} \& {Weinberg}(1996)}]{1996ApJ...468..462G}
{Gould}, A., \& {Weinberg}, D.~H. 1996, \apj, 468, 462

\bibitem[{{Guo} {et~al.}(2011){Guo}, {White}, {Boylan-Kolchin}, {De Lucia},
  {Kauffmann}, {Lemson}, {Li}, {Springel}, \& {Weinmann}}]{2011MNRAS.413..101G}
{Guo}, Q., {White}, S., {Boylan-Kolchin}, M., {et~al.} 2011, \mnras, 413, 101

\bibitem[{{Haardt} \& {Madau}(2012)}]{2012ApJ...746..125H}
{Haardt}, F., \& {Madau}, P. 2012, \apj, 746, 125

\bibitem[{{Haiman} {et~al.}(2000){Haiman}, {Spaans}, \&
  {Quataert}}]{2000ApJ...537L...5H}
{Haiman}, Z., {Spaans}, M., \& {Quataert}, E. 2000, \apjl, 537, L5

\bibitem[{{Hayes} {et~al.}(2011){Hayes}, {Schaerer}, {{\"O}stlin}, {Mas-Hesse},
  {Atek}, \& {Kunth}}]{2011ApJ...730....8H}
{Hayes}, M., {Schaerer}, D., {{\"O}stlin}, G., {et~al.} 2011, \apj, 730, 8

\bibitem[{{Iye} {et~al.}(2006){Iye}, {Ota}, {Kashikawa}, {Furusawa},
  {Hashimoto}, {Hattori}, {Matsuda}, {Morokuma}, {Ouchi}, \&
  {Shimasaku}}]{2006Natur.443..186I}
{Iye}, M., {Ota}, K., {Kashikawa}, N., {et~al.} 2006, \nat, 443, 186

\bibitem[{{Jensen} {et~al.}(2012){Jensen}, {Laursen}, {Mellema}, {Iliev},
  {Sommer-Larsen}, \& {Shapiro}}]{2012arXiv1206.4028J}
{Jensen}, H., {Laursen}, P., {Mellema}, G., {et~al.} 2012, ArXiv e-prints

\bibitem[{{Jiang} {et~al.}(2011){Jiang}, {Egami}, {Kashikawa}, {Walth},
  {Matsuda}, {Shimasaku}, {Nagao}, {Ota}, \& {Ouchi}}]{2011ApJ...743...65J}
{Jiang}, L., {Egami}, E., {Kashikawa}, N., {et~al.} 2011, \apj, 743, 65

\bibitem[{{Karzas} \& {Latter}(1961)}]{1961ApJS....6..167K}
{Karzas}, W.~J., \& {Latter}, R. 1961, \apjs, 6, 167

\bibitem[{{Kashikawa} {et~al.}(2006){Kashikawa}, {Shimasaku}, {Malkan}, {Doi},
  {Matsuda}, {Ouchi}, {Taniguchi}, {Ly}, {Nagao}, {Iye}, {Motohara},
  {Murayama}, {Murozono}, {Nariai}, {Ohta}, {Okamura}, {Sasaki}, {Shioya}, \&
  {Umemura}}]{2006ApJ...648....7K}
{Kashikawa}, N., {Shimasaku}, K., {Malkan}, M.~A., {et~al.} 2006, \apj, 648, 7

\bibitem[{{Kennicutt}(1998{\natexlab{a}})}]{1998ARA&A..36..189K}
{Kennicutt}, Jr., R.~C. 1998{\natexlab{a}}, \araa, 36, 189

\bibitem[{{Kennicutt}(1998{\natexlab{b}})}]{1998ApJ...498..541K}
---. 1998{\natexlab{b}}, \apj, 498, 541

\bibitem[{{Kennicutt} {et~al.}(1994){Kennicutt}, {Tamblyn}, \&
  {Congdon}}]{1994ApJ...435...22K}
{Kennicutt}, Jr., R.~C., {Tamblyn}, P., \& {Congdon}, C.~E. 1994, \apj, 435, 22

\bibitem[{{Komatsu} {et~al.}(2011){Komatsu}, {Smith}, {Dunkley}, {Bennett},
  {Gold}, {Hinshaw}, {Jarosik}, {Larson}, {Nolta}, {Page}, {Spergel},
  {Halpern}, {Hill}, {Kogut}, {Limon}, {Meyer}, {Odegard}, {Tucker}, {Weiland},
  {Wollack}, \& {Wright}}]{2011ApJS..192...18K}
{Komatsu}, E., {Smith}, K.~M., {Dunkley}, J., {et~al.} 2011, \apjs, 192, 18

\bibitem[{{Larson} {et~al.}(2011){Larson}, {Dunkley}, {Hinshaw}, {Komatsu},
  {Nolta}, {Bennett}, {Gold}, {Halpern}, {Hill}, {Jarosik}, {Kogut}, {Limon},
  {Meyer}, {Odegard}, {Page}, {Smith}, {Spergel}, {Tucker}, {Weiland},
  {Wollack}, \& {Wright}}]{2011ApJS..192...16L}
{Larson}, D., {Dunkley}, J., {Hinshaw}, G., {et~al.} 2011, \apjs, 192, 16

\bibitem[{{Latif} {et~al.}(2011){Latif}, {Schleicher}, {Spaans}, \&
  {Zaroubi}}]{2011MNRAS.413L..33L}
{Latif}, M.~A., {Schleicher}, D.~R.~G., {Spaans}, M., \& {Zaroubi}, S. 2011,
  \mnras, 413, L33

\bibitem[{{Lehnert} {et~al.}(2010){Lehnert}, {Nesvadba}, {Cuby}, {Swinbank},
  {Morris}, {Cl{\'e}ment}, {Evans}, {Bremer}, \& {Basa}}]{2010Natur.467..940L}
{Lehnert}, M.~D., {Nesvadba}, N.~P.~H., {Cuby}, J.-G., {et~al.} 2010, \nat,
  467, 940

\bibitem[{{Lidz} {et~al.}(2011){Lidz}, {Furlanetto}, {Oh}, {Aguirre}, {Chang},
  {Dor{\'e}}, \& {Pritchard}}]{2011ApJ...741...70L}
{Lidz}, A., {Furlanetto}, S.~R., {Oh}, S.~P., {et~al.} 2011, \apj, 741, 70

\bibitem[{{Madau} \& {Pozzetti}(2000)}]{2000MNRAS.312L...9M}
{Madau}, P., \& {Pozzetti}, L. 2000, \mnras, 312, L9

\bibitem[{{Maraston}(2005)}]{2005MNRAS.362..799M}
{Maraston}, C. 2005, \mnras, 362, 799

\bibitem[{{Martin}(1988)}]{1988ApJS...66..125M}
{Martin}, P.~G. 1988, \apjs, 66, 125

\bibitem[{{Mesinger} \& {Furlanetto}(2007)}]{2007ApJ...669..663M}
{Mesinger}, A., \& {Furlanetto}, S. 2007, \apj, 669, 663

\bibitem[{{Mesinger} \& {Haiman}(2007)}]{2007ApJ...660..923M}
{Mesinger}, A., \& {Haiman}, Z. 2007, \apj, 660, 923

\bibitem[{{Mitra} {et~al.}(2012){Mitra}, {Choudhury}, \&
  {Ferrara}}]{2012MNRAS.419.1480M}
{Mitra}, S., {Choudhury}, T.~R., \& {Ferrara}, A. 2012, \mnras, 419, 1480

\bibitem[{{Oesch} {et~al.}(2012){Oesch}, {Bouwens}, {Illingworth}, {Labb{\'e}},
  {Trenti}, {Gonzalez}, {Carollo}, {Franx}, {van Dokkum}, \&
  {Magee}}]{2012ApJ...745..110O}
{Oesch}, P.~A., {Bouwens}, R.~J., {Illingworth}, G.~D., {et~al.} 2012, \apj,
  745, 110

\bibitem[{{Ota} {et~al.}(2008){Ota}, {Iye}, {Kashikawa}, {Shimasaku},
  {Kobayashi}, {Totani}, {Nagashima}, {Morokuma}, {Furusawa}, {Hattori},
  {Matsuda}, {Hashimoto}, \& {Ouchi}}]{2008ApJ...677...12O}
{Ota}, K., {Iye}, M., {Kashikawa}, N., {et~al.} 2008, \apj, 677, 12

\bibitem[{{Ota} {et~al.}(2010){Ota}, {Iye}, {Kashikawa}, {Shimasaku}, {Ouchi},
  {Totani}, {Kobayashi}, {Nagashima}, {Harayama}, {Kodaka}, {Morokuma},
  {Furusawa}, {Tajitsu}, \& {Hattori}}]{2010ApJ...722..803O}
---. 2010, \apj, 722, 803

\bibitem[{{Ouchi} {et~al.}(2008){Ouchi}, {Shimasaku}, {Akiyama}, {Simpson},
  {Saito}, {Ueda}, {Furusawa}, {Sekiguchi}, {Yamada}, {Kodama}, {Kashikawa},
  {Okamura}, {Iye}, {Takata}, {Yoshida}, \& {Yoshida}}]{2008ApJS..176..301O}
{Ouchi}, M., {Shimasaku}, K., {Akiyama}, M., {et~al.} 2008, \apjs, 176, 301

\bibitem[{{Ouchi} {et~al.}(2010){Ouchi}, {Shimasaku}, {Furusawa}, {Saito},
  {Yoshida}, {Akiyama}, {Ono}, {Yamada}, {Ota}, {Kashikawa}, {Iye}, {Kodama},
  {Okamura}, {Simpson}, \& {Yoshida}}]{2010ApJ...723..869O}
{Ouchi}, M., {Shimasaku}, K., {Furusawa}, H., {et~al.} 2010, \apj, 723, 869

\bibitem[{{Pawlik} {et~al.}(2010){Pawlik}, {Schaye}, \& {van
  Scherpenzeel}}]{2010ASPC..432..230P}
{Pawlik}, A.~H., {Schaye}, J., \& {van Scherpenzeel}, E. 2010, in Astronomical
  Society of the Pacific Conference Series, Vol. 432, New Horizons in
  Astronomy: Frank N. Bash Symposium 2009, ed. {L.~M.~Stanford, J.~D.~Green,
  L.~Hao, \& Y.~Mao}, 230

\bibitem[{{Pengelly}(1964)}]{1964MNRAS.127..145P}
{Pengelly}, R.~M. 1964, \mnras, 127, 145

\bibitem[{{Popesso} {et~al.}(2012){Popesso}, {Biviano}, {Rodighiero},
  {Baronchelli}, {Salvato}, {Saintonge}, {Finoguenov}, {Magnelli}, {Gruppioni},
  {Pozzi}, {Lutz}, {Elbaz}, {Altieri}, {Andreani}, {Aussel}, {Berta}, {Capak},
  {Cava}, {Cimatti}, {Coia}, {Daddi}, {Dannerbauer}, {Dickinson}, {Dasyra},
  {Fadda}, {F{\"o}rster Schreiber}, {Genzel}, {Hwang}, {Kartaltepe}, {Ilbert},
  {Le Floch}, {Leiton}, {Magdis}, {Nordon}, {Patel}, {Poglitsch}, {Riguccini},
  {Sanchez Portal}, {Shao}, {Tacconi}, {Tomczak}, {Tran}, \&
  {Valtchanov}}]{2012A&A...537A..58P}
{Popesso}, P., {Biviano}, A., {Rodighiero}, G., {et~al.} 2012, \aap, 537, A58

\bibitem[{{Razoumov} \& {Sommer-Larsen}(2010)}]{2010ApJ...710.1239R}
{Razoumov}, A.~O., \& {Sommer-Larsen}, J. 2010, \apj, 710, 1239

\bibitem[{{Salvaterra} {et~al.}(2009){Salvaterra}, {Della Valle}, {Campana},
  {Chincarini}, {Covino}, {D'Avanzo}, {Fern{\'a}ndez-Soto}, {Guidorzi},
  {Mannucci}, {Margutti}, {Th{\"o}ne}, {Antonelli}, {Barthelmy}, {de Pasquale},
  {D'Elia}, {Fiore}, {Fugazza}, {Hunt}, {Maiorano}, {Marinoni}, {Marshall},
  {Molinari}, {Nousek}, {Pian}, {Racusin}, {Stella}, {Amati}, {Andreuzzi},
  {Cusumano}, {Fenimore}, {Ferrero}, {Giommi}, {Guetta}, {Holland}, {Hurley},
  {Israel}, {Mao}, {Markwardt}, {Masetti}, {Pagani}, {Palazzi}, {Palmer},
  {Piranomonte}, {Tagliaferri}, \& {Testa}}]{2009Natur.461.1258S}
{Salvaterra}, R., {Della Valle}, M., {Campana}, S., {et~al.} 2009, \nat, 461,
  1258

\bibitem[{{Santos} {et~al.}(2010){Santos}, {Ferramacho}, {Silva}, {Amblard}, \&
  {Cooray}}]{2010MNRAS.406.2421S}
{Santos}, M.~G., {Ferramacho}, L., {Silva}, M.~B., {Amblard}, A., \& {Cooray},
  A. 2010, \mnras, 406, 2421

\bibitem[{{Santos} {et~al.}(2011){Santos}, {Silva}, {Pritchard}, {Cen}, \&
  {Cooray}}]{2011A&A...527A..93S}
{Santos}, M.~G., {Silva}, M.~B., {Pritchard}, J.~R., {Cen}, R., \& {Cooray}, A.
  2011, \aap, 527, A93

\bibitem[{{Santos}(2004)}]{2004MNRAS.349.1137S}
{Santos}, M.~R. 2004, \mnras, 349, 1137

\bibitem[{{Schaerer}(2002)}]{2002A&A...382...28S}
{Schaerer}, D. 2002, \aap, 382, 28

\bibitem[{{Sheth} \& {Tormen}(1999)}]{1999MNRAS.308..119S}
{Sheth}, R.~K., \& {Tormen}, G. 1999, \mnras, 308, 119

\bibitem[{{Shibuya} {et~al.}(2011){Shibuya}, {Kashikawa}, {Ota}, {Iye},
  {Ouchi}, {Furusawa}, {Shimasaku}, \& {Hattori}}]{2011arXiv1112.3997S}
{Shibuya}, T., {Kashikawa}, N., {Ota}, K., {et~al.} 2011, ArXiv e-prints

\bibitem[{{Shimasaku} {et~al.}(2006){Shimasaku}, {Kashikawa}, {Doi}, {Ly},
  {Malkan}, {Matsuda}, {Ouchi}, {Hayashino}, {Iye}, {Motohara}, {Murayama},
  {Nagao}, {Ohta}, {Okamura}, {Sasaki}, {Shioya}, \&
  {Taniguchi}}]{2006PASJ...58..313S}
{Shimasaku}, K., {Kashikawa}, N., {Doi}, M., {et~al.} 2006, \pasj, 58, 313

\bibitem[{{Shull} \& {van Steenberg}(1985)}]{1985ApJ...298..268S}
{Shull}, J.~M., \& {van Steenberg}, M.~E. 1985, \apj, 298, 268

\bibitem[{{Shull} {et~al.}(2011){Shull}, {Harness}, {Trenti}, \&
  {Smith}}]{2011arXiv1108.3334S}
{Shull}, M., {Harness}, A., {Trenti}, M., \& {Smith}, B. 2011, ArXiv e-prints

\bibitem[{{Siana} {et~al.}(2007){Siana}, {Teplitz}, {Colbert}, {Ferguson},
  {Dickinson}, {Brown}, {Conselice}, {de Mello}, {Gardner}, {Giavalisco}, \&
  {Menanteau}}]{2007ApJ...668...62S}
{Siana}, B., {Teplitz}, H.~I., {Colbert}, J., {et~al.} 2007, \apj, 668, 62

\bibitem[{{Smith} {et~al.}(2011){Smith}, {Hallman}, {Shull}, \&
  {O'Shea}}]{2011ApJ...731....6S}
{Smith}, B.~D., {Hallman}, E.~J., {Shull}, J.~M., \& {O'Shea}, B.~W. 2011,
  \apj, 731, 6

\bibitem[{{Springel} {et~al.}(2005){Springel}, {White}, {Jenkins}, {Frenk},
  {Yoshida}, {Gao}, {Navarro}, {Thacker}, {Croton}, {Helly}, {Peacock}, {Cole},
  {Thomas}, {Couchman}, {Evrard}, {Colberg}, \& {Pearce}}]{2005Natur.435..629S}
{Springel}, V., {White}, S.~D.~M., {Jenkins}, A., {et~al.} 2005, \nat, 435, 629

\bibitem[{{Steidel} {et~al.}(2011){Steidel}, {Bogosavljevi{\'c}}, {Shapley},
  {Kollmeier}, {Reddy}, {Erb}, \& {Pettini}}]{2011ApJ...736..160S}
{Steidel}, C.~C., {Bogosavljevi{\'c}}, M., {Shapley}, A.~E., {et~al.} 2011,
  \apj, 736, 160

\bibitem[{{Taniguchi} {et~al.}(2005){Taniguchi}, {Ajiki}, {Nagao}, {Shioya},
  {Murayama}, {Kashikawa}, {Kodaira}, {Kaifu}, {Ando}, {Karoji}, {Akiyama},
  {Aoki}, {Doi}, {Fujita}, {Furusawa}, {Hayashino}, {Iwamuro}, {Iye},
  {Kobayashi}, {Kodama}, {Komiyama}, {Matsuda}, {Miyazaki}, {Mizumoto},
  {Morokuma}, {Motohara}, {Nariai}, {Ohta}, {Ohyama}, {Okamura}, {Ouchi},
  {Sasaki}, {Sato}, {Sekiguchi}, {Shimasaku}, {Tamura}, {Umemura}, {Yamada},
  {Yasuda}, \& {Yoshida}}]{2005PASJ...57..165T}
{Taniguchi}, Y., {Ajiki}, M., {Nagao}, T., {et~al.} 2005, \pasj, 57, 165

\bibitem[{{Visbal} \& {Loeb}(2010)}]{2010JCAP...11..016V}
{Visbal}, E., \& {Loeb}, A. 2010, JCAP, 11, 16

\bibitem[{{Wang} {et~al.}(2006){Wang}, {Tegmark}, {Santos}, \&
  {Knox}}]{2006ApJ...650..529W}
{Wang}, X., {Tegmark}, M., {Santos}, M.~G., \& {Knox}, L. 2006, \apj, 650, 529

\bibitem[{{Wise} \& {Cen}(2009)}]{2009ApJ...693..984W}
{Wise}, J.~H., \& {Cen}, R. 2009, \apj, 693, 984

\bibitem[{{Wyithe} {et~al.}(2007){Wyithe}, {Loeb}, \&
  {Schmidt}}]{2007arXiv0705.1825W}
{Wyithe}, S., {Loeb}, A., \& {Schmidt}, B. 2007, ArXiv e-prints

\bibitem[{{Yajima} {et~al.}(2011){Yajima}, {Choi}, \&
  {Nagamine}}]{2011MNRAS.412..411Y}
{Yajima}, H., {Choi}, J.-H., \& {Nagamine}, K. 2011, \mnras, 412, 411

\bibitem[{{Zahn} {et~al.}(2011){Zahn}, {Reichardt}, {Shaw}, {Lidz}, {Aird},
  {Benson}, {Bleem}, {Carlstrom}, {Chang}, {Cho}, {Crawford}, {Crites}, {de
  Haan}, {Dobbs}, {Dore}, {Dudley}, {George}, {Halverson}, {Holder},
  {Holzapfel}, {Hoover}, {Hou}, {Hrubes}, {Joy}, {Keisler}, {Knox}, {Lee},
  {Leitch}, {Lueker}, {Luong-Van}, {McMahon}, {Mehl}, {Meyer}, {Millea},
  {Mohr}, {Montroy}, {Natoli}, {Padin}, {Plagge}, {Pryke}, {Ruhl}, {Schaffer},
  {Shirokoff}, {Spieler}, {Staniszewski}, {Stark}, {Story}, {van Engelen},
  {Vanderlinde}, {Vieira}, \& {Williamson}}]{2011arXiv1111.6386Z}
{Zahn}, O., {Reichardt}, C.~L., {Shaw}, L., {et~al.} 2011, ArXiv e-prints

\bibitem[{{Zaldarriaga} {et~al.}(2008){Zaldarriaga}, {Colombo}, {Komatsu},
  {Lidz}, {Mortonson}, {Oh}, {Pierpaoli}, {Verde}, \&
  {Zahn}}]{2008arXiv0811.3918Z}
{Zaldarriaga}, M., {Colombo}, L., {Komatsu}, E., {et~al.} 2008, ArXiv e-prints

\bibitem[{{Zheng} {et~al.}(2010){Zheng}, {Cen}, {Trac}, \&
  {Miralda-Escud{\'e}}}]{2010ApJ...716..574Z}
{Zheng}, Z., {Cen}, R., {Trac}, H., \& {Miralda-Escud{\'e}}, J. 2010, \apj,
  716, 574

\bibitem[{{Zheng} {et~al.}(2011){Zheng}, {Cen}, {Weinberg}, {Trac}, \&
  {Miralda-Escud{\'e}}}]{2011ApJ...739...62Z}
{Zheng}, Z., {Cen}, R., {Weinberg}, D., {Trac}, H., \& {Miralda-Escud{\'e}}, J.
  2011, \apj, 739, 62

\end{thebibliography}

\end{document}